\documentclass[12pt]{article}

\pdfoutput=1

\usepackage{graphics}
\usepackage{epsfig}

\usepackage{amsfonts}
\usepackage{amssymb}
\usepackage{amsmath}
\usepackage{dsfont}
\usepackage{multirow}
\usepackage{latexsym}
\usepackage{verbatim}
\usepackage{mcite}
\usepackage{cite}
\usepackage{units}
\usepackage[all]{xy}
\usepackage[usenames,dvipsnames,table]{xcolor}
\usepackage{cancel}
\usepackage{slashed}
\usepackage{hyperref}
\usepackage{physics}
\usepackage{breqn}

\usepackage{tikz}
\usetikzlibrary{shapes,arrows,positioning,snakes}

\numberwithin{equation}{section}

\def\beq{\begin{equation}}
\def\eeq{\end{equation}}
\def\beqa{\begin{eqnarray}}
\def\eeqa{\end{eqnarray}}
\newcommand{\f}{\frac}

\def\bfone{\relax{\rm 1\kern-.35em 1}}

\def\tsum{\textstyle \sum}

\def\vphi{{\vec \phi}}
\def\vpi{{\vec \pi}}

\def\vr{{\mathbf{r}}}
\def\vx{{\mathbf{x}}}

\def\vp{{\mathbf{p}}}
\def\vq{{\mathbf{q}}}
\def\vk{{\mathbf{k}}}

\def\vx{{\mathbf{x}}}

\newcommand{\cB}{{\cal B}}
\newcommand{\cC}{{\cal C}}
\newcommand{\cD}{{\cal D}}

\newcommand{\cH}{{\cal H}}
\newcommand{\cK}{{\cal K}}

\newcommand{\cN}{{\cal N}}
\newcommand{\cP}{{\cal P}}
\newcommand{\cO}{{\cal O}}

\newcommand{\cU}{{\cal U}}

\newcommand{\bbR}{{\mathbb{R}}}

\newcommand{\be}{\begin{equation}}
\newcommand{\ee}{\end{equation}}
\newcommand{\ben}{\begin{displaymath}}
\newcommand{\een}{\end{displaymath}}
\newcommand{\bea}{\begin{eqnarray}}
\newcommand{\eea}{\end{eqnarray}}

\newcommand{\bean}{\begin{eqnarray*}}
\newcommand{\eean}{\end{eqnarray*}}

\newcommand{\bsPhi}{\boldsymbol \Phi}
\newcommand{\bsPi}{\boldsymbol \Pi}

\DeclareMathAlphabet{\mathpzc}{OT1}{pzc}{m}{it}

\begin{document}

\pagestyle{plain}

\begin{titlepage}
\begin{flushright}
\end{flushright}

\bigskip

    \begin{center}
		{\Large \bf{The Large $N$ Limit of icMERA and Holography
	}}
	
	\vskip 1.5cm

	
	
{\bf Jos\'e J. Fern\'andez-Melgarejo$^\spadesuit$\footnote{melgarejo@at@um.es} and  Javier Molina-Vilaplana$^\blacklozenge$\footnote{javi.molina@at@upct.es}}

\begin{center}
${}^\spadesuit${\it Departamento de F\'isica, Universidad de Murcia,}\\
{\it Campus de Espinardo, 30100 Murcia, Spain}

\vspace*{0.5cm}

${}^\blacklozenge${\it Departamento de Autom\'atica, Universidad Polit\'ecnica de Cartagena,}\\
{\it Calle Dr. Fleming, S/N, 30202 Cartagena, Spain}
\end{center}

\vspace{1cm}

	\today
	
\vspace{1cm}

    \end{center}

\begin{abstract}
In this work, we compute the entanglement entropy in continuous icMERA tensor networks for large $N$ models at strong coupling. Our results show that the $1/N$ quantum corrections to the Fisher information metric (interpreted as a local bond dimension of the tensor network) in an icMERA circuit, are related to quantum corrections to the minimal area surface in the Ryu-Takayanagi formula. 
Upon picking two different non-Gaussian entanglers to build the icMERA circuit, the results for the entanglement entropy only differ at subleading orders in $1/G_N$, \emph{i.e.}, at the structure of the quantum corrections in the bulk. The fact that the large $N$ part of the entropy can be always related to the leading area term of the holographic calculation is very suggestive. These results, constitute the first tensor network calculations at large $N$ and strong coupling simultaneously, pushing the field of tensor network descriptions of the emergence of dual spacetime geometries from the structure of entanglement in quantum field theory.
\end{abstract}

\end{titlepage}

\tableofcontents

\section{Introduction and Results}
\label{sec:intro}

\subsubsection*{Motivation}

Since the pioneering work of Swingle \cite{Swingle:2009bg} and despite the substantial developments done in \cite{Swingle:2012wq,Nozaki:2012zj,Mollabashi:2013lya, Beny:2011vh, Miyaji:2014mca, Molina-Vilaplana:2015mja, Molina-Vilaplana:2015rra, Czech:2015kbp, Czech:2015qta, Bao:2015uaa, Miyaji:2015fia, Miyaji:2015yva, Hayden:2016cfa, Caputa:2017urj, Caputa:2017yrh,Fliss:2016ifp}, a more precise connection between entanglement renormalization tensor networks (MERA) \cite{Vidal:2007hda, Haegeman:2011uy} and AdS/CFT \cite{Maldacena:1997re, Gubser:1998bc, Witten:1998qj} has not been fully established. The lack of a MERA circuit formulation where to consistently take the large $N$ limit at the strong coupling regime, which are the fundamental tenets of holography, hinders this potential correspondence. Upon conceding the pertinence of such connection, the existence of a precise setting explaining how entanglement renormalization may describe the relationship between geometry and entanglement in actual AdS/CFT, would shed light on the Hilbert space microstructure of holographic spacetimes. In addition, this would provide us a guide to investigate the holographic duality away from large $N$ and strong coupling limits.

\medskip

A consistent AdS/MERA correspondence also requires to clarify how a tensor network directly describes sub-AdS scale physics \cite{Bao:2015uaa}.  Without the addition of supplementary structure to the standard MERA tensors, this is not possible and thus the feasibility for the discrete MERA to be a complete description of the gravity theory is limited. This problem has been partially explored in \cite{Bao:2018pvs, Bao:2019fpq}. 

Another approach consists of using generalizations of the continuous MERA (cMERA) with some kind of local expansions of the tensor structure of the circuit that would be able to give account of sub-AdS physics. In this sense, a new class of non-Gaussian cMERA tensor network (icMERA) has been recently introduced to deal with both weak \cite{Cotler:2018ehb,Cotler:2018ufx} and strongly interacting field theories \cite{Fernandez-Melgarejo:2019sjo, Fernandez-Melgarejo:2020fzw}.

\medskip

At infinite $N$, the holographic dictionary maps quantum {\it entanglement} in a region $A$ of a field theory to the classical area of a minimal surface $\gamma_A$ in its gravitational dual \cite{Ryu:2006bv} throughout the Ryu-Takayanagi formula:
\begin{align}
S_A = \frac{{\rm Area}(\gamma_A)}{4 G_N}   \,.
\label{eq:rt1}
\end{align}
If we consider now a $1/N$ expansion, the gravitational theory is no longer strictly classical, and the first quantum correction to equation \eqref{eq:rt1} in the bulk has been argued to be \cite{Faulkner:2013ana}:
\begin{align}
S_A = \frac{{\rm Area}(\gamma_A)}{4 G_N} + S_{\rm q} \,,
\label{eq:flm}
\end{align}
where $S_{\rm q}$ accounts for the first quantum corrections in the bulk. These quantum corrections are comprised into the \emph{bulk} entanglement entropy, \emph{i.e.}, the entanglement entropy of the low energy degrees of freedom  of an effective quantum field theory in the bulk, plus other terms such as corrections to the area of the minimal surface due to the one loop change in the bulk metric. As a result, at finite $N$, the boundary entanglement is given in terms of a leading order classical area contribution plus a subleading entanglement term.  The equation \eqref{eq:flm}  thus determines to what extent quantum degrees of freedom in the bulk are encoded into the boundary. Despite the conceptual importance of  \eqref{eq:flm},  obtaining explicit computations of these quantum corrections turns out to be a rather difficult task in general. For instance, it is hard to calculate the bulk entanglement entropy from a straight bulk computation.

\medskip

It is thus our aim in this work to explore the large $N$ limit of strongly coupled theories through icMERA to investigate  possible connections with holography. More concretely, we are interested in thrusting this correspondence through explicit quantitative calculations of quantities such as entanglement entropy for which, a well established holographic correspondence is given in terms of the Ryu-Takayanagi formula. 

\subsubsection*{Results}

In this work, we study the structure of entanglement entropy in cMERA tensor networks for $O(N)$ vector models at strong coupling and calculate their large $N$ limit. Here, it is worth to mention \cite{Fliss:2016ifp}, where authors studied the exact renormalization group (ERG, \cite{Polchinski:1983gv}) for free $O(N)$ vector models (at large $N$) by extending the ERG formalism to wavefunctionals. In such formalism, the ERG flow of the ground state and some class of excited states is implemented in terms of local unitary operators. As a consequence, the ERG equations can be interpreted in terms of a continuous tensor network that shares many general features with a Gaussian cMERA but differs in important ways. 

With this, upon studying the single scalar case in Section \ref{sec:scalar} we have laid down the foundations of the calculation of entanglement entropy using truly non-Gaussian cMERA circuits. In particular, we have obtained the half space entanglement entropy for a single scalar theory with arbitrary potential using various icMERA entanglers, $B\sim \pi\phi^2$, Eq. \eqref{eq:B-piphi2} and $B\sim \pi\phi^3$, Eq. \eqref{eq:B-piphi3}. In both cases, the result exhibits the following structure:
\begin{align}
\tilde S_A = S^{(0)}_A + s^2\, S^{\chi,\zeta}_A + \cO(s^4)\, ,
\end{align}
where
$S_A^{(0)}$ is the Gaussian area-law term and $S_A^{\chi,\zeta}$ consists of the non-Gaussian contributions (\emph{c.f.} \eqref{eq:ee_piphi2} and \eqref{eq:ee_piphi3}). $s$ is the variational parameter which, upon minimization is proportial to $\lambda$.
\footnote{We remark that, despite $|s|\ll 1$, $\lambda$ is not necessarily perturbative.}

Regarding the $O(N)$ vector model, our primary goal is to provide some quantitative salient features of the correspondence at large $N$ and strong coupling. To do so, an important caveat is worth to be mentioned here. In Section \ref{sec:ON-icmera} we have formulated the $O(N)$ free cMERA and icMERA circuit from first principles, where two non-Gaussian entanglers, $B_N\sim (\bsPi\cdot\bsPhi)( \bsPhi\cdot\bsPhi)$, \eqref{eq:entangler-ON-G-sym}, and $\cB_N\sim \pi_R\ \phi_T\cdot\phi_T$ \eqref{eq:entangler-ON-NG} have been introduced.

Then, using the properties of the trial wavefunctionals generated by the $O(N)$ icMERA circuit, we calculate the half space entanglement entropy by using the same techniques as in \cite{Fernandez-Melgarejo:2020utg,Fernandez-Melgarejo:2021ymz}.  When considering a $1/N$ expansion, the entanglement entropy for $\cB_N$ is given by (see \eqref{eq:largeN_ee_chi})
\begin{align}
S_A 
=&\ 
S_{T}^{(N)} + \tilde S_{R,\chi}^{(1)} 
\, ,
\end{align}
where $S_T^{(N)}\propto N$ is the leading term associated to the Gaussian contributions of the fields and   $\tilde S_{R,\chi}^{(1)}$, which is $\cO(N^0)$, accounts for the non-Gaussian contributions generated by the icMERA circuit.

Similarly, when implementing icMERA with the entangler $B_N$ in \eqref{eq:entangler-ON-G-sym}, the resulting half space entanglement entropy is given by (see \eqref{eq:ON-ee-largeN1})
\begin{align}
S_A
=&\ S^{(N)} + S^{(1)}_{\zeta}
\, ,
\end{align}
where $S^{(N)}\propto N$ is again the leading term associated to the Gaussian variational ansatz and $S^{(1)}_{\zeta}$, which is $\cO(1)$, is entirely generated by the entangler $B_N$.

From these results, we conclude that the leading $1/N$ quantum corrections to the Fisher information metric in an icMERA circuit potentially reproduce the terms associated to the quantum corrections of the holographic entanglement entropy. That is to say, in our construction, these $1/N$ corrections to the icMERA Fisher metric can be explicitly related to the quantum corrections to the RT semiclassical leading term as follows (see \eqref{eq:entropy_qcorr_icmera} for further details):
\begin{align}
S_{\textrm q} 
\sim 
\expval{\hat B_N(u)^2}_\Omega^{1/2}
 \, .
\end{align}
where $\ket\Omega$ is the purely disentangled Gaussian reference state in the usual cMERA formalism. Similar ideas have been explored in \cite{Banerjee:2017qti} where arguments relating the reduced Fisher information to the {\it canonical energy} have been used. 

In addition to this, from Sections \ref{sec:scalar} and \ref{sec:ON-ee}, we conclude that in cMERA the Fisher information metric turns out to be essential to calculate the entanglement entropy,  as it is the quantity that naturally plays the r\^ole of the bond dimension in the continuous version of the tensor network.
Given the analogy of the Fisher metric as a local bond dimension in cMERA,  we show that adding more entanglers, and consequently having more terms in the sum, is equivalent to increasing the bond dimension in the discrete MERA circuit. 

Precisely, in \cite{Bao:2018pvs, Bao:2019fpq} authors gave a general procedure for constructing tensor networks for geometric states in the AdS/CFT. Taking  formula \eqref{eq:flm}, they noted that in any such tensor network, the bond dimensions must be determined by the areas of a corresponding extremal surface in the bulk $-\, $ in our language, the leading term in the cMERA density of disentanglers must give account for the term ${\rm Area}/4 G_N\,-$, while the subleading terms in the bond dimension correspond to fluctuations in the areas of those extremal surfaces due to, \emph{e.g.}, graviton effects. Consequently, our results, which are consistent with these works, may help to understand the Hilbert space structure of an underlying dual gravitational description of the tensor network and thus to firmly establish a duality between cMERA and holography.

In order to sustain our calculation of entanglement entropy based on a Fisher metric-inspired definition of a bond dimension, we have obtained the entanglement entropy by means of a parallel computation based solely on the Gaussian nature of the icMERA wavefunctional in the basis of non-linearly deformed fields which the method is based on and the results for the half space entropy in a Gaussian cMERA \cite{Fernandez-Melgarejo:2021ymz}.

As a result, our approach to address the question of the large $N$ and strong coupling in continuous tensor networks exhibits the natural emergence of a non-fluctuating smooth geometry description, which represents a crucial feature of the holographic duality. Namely, our holographic interpretation of the large $N$ limit of icMERA tensor networks imply that any two choices of the non-Gaussian part of the circuit generator yield results for the entanglement that differ only at subleading orders in $1/G_N$; that is to say, at the structure of the quantum corrections. Thus, in the large $N$ limit, the predictions for any two tensor networks corresponding to different icMERA schemes should converge up to subleading corrections. The fact that the large $N$ part of the entropy can always be related to the leading area term of the holographic calculation is thus very suggestive. We remark here that we have only used tensor network \emph{technology} (\emph{i.e.}, quantum information properties of the boundary state) to derive holographic spatial geometries consistent with the structure of quantum entanglement in field theory. In other words, our icMERA analysis in the large $N$ and strong coupling limit shows, provides a detailed and concrete setting showing how continuous tensor networks can been used to describe dual spacetime geometries in terms of the structure of entanglement in QFT states.
 
\medskip 
 
Let us finally comment on the structure of this work. In Section \ref{sec:scalar} we provide the basic conceptual structure of the paper by working with a single self-interacting scalar theory. We introduce icMERA circuits and compute the half space entanglement entropy of this theory. As the icMERA circuit allows us to address the non-perturbative regime, the results are eloquent enough to illustrate the behavior of entanglement at strong coupling, one of the key ingredients in order to tackle the connection with holography. The natural subsequent step is dealing both with strong coupling and a large number of fields. To do this, we introduce in Section \ref{sec:ON-icmera}, a novel class of icMERA circuits devised to deal with interacting vector models. Then in Section \ref{sec:ON-ee}, we obtain the icMERA entanglement entropy of an $O(N)$ interacting vector model. As commented above, two parallel computations are carried out in order to firmly check that a computation through a properly defined bond dimension is consistent. At the end of the Section, the large $N$ limit of the results are taken, allowing for a holographic discussion of the results that is performed in Section \ref{sec:holo_icmera}. Finally we conclude by giving some briefs prospects on our work.

\section{Single Scalar Theories}
\label{sec:scalar}

In this section we will consider the so called icMERA circuit for any theory with a single real scalar field. It consists of an entanglement renormalization method that generates non-Gaussian states, which are proven to be useful to obtain an approximated ground state. Obviously, because the trial states are genuinely non-Gaussian, this is a suitable tensor network for approaching the ground state of interacting scalar theories. 

The icMERA circuit \cite{Fernandez-Melgarejo:2019sjo,Fernandez-Melgarejo:2020fzw} can be used to inspect the ground state of any interacting theory with a (non-)polynomial potential. In this method, the specific choice of the scalar potential will only determine the value of the circuit variational parameters upon the minimization of the energy functional. More specifically, while the functional structure of any $n$-point correlator is not affected by the chosen interacting model, its value will be entirely shaped by the optimized variational parameters (and hence, by the specific potential).

After introducing icMERA, following \cite{Fernandez-Melgarejo:2021ymz}, in this section we will calculate the half space entanglement entropy in terms of the variational parameters of the tensor network. In particular, we will show that for our trial states, a quantity that may be interpreted as the bond dimension in the discrete versions of MERA, is exactly given by the Fisher information metric of the icMERA circuit.

To fix our notation, let us firstly introduce the free theory. The Lagrangian density is 
\begin{align}
\mathcal{L} = \frac{1}{2}\left[(\partial \phi)^2 - m^2 \phi^2\right]\, ,
\end{align}
with the field operator satisfying commutation relations 
\begin{align}
[\phi(\vp),\pi(\vq)]=i\delta(\vp+\vq)
\ ,
\qquad
[\phi(\vp),\phi(\vq)]=[\pi(\vp),\pi(\vq)]=0
\ .
\end{align}

\subsection{An icMERA Primer}
\label{sec:scalar-icmera}

In \cite{Fernandez-Melgarejo:2019sjo,Fernandez-Melgarejo:2020fzw} a non-Gaussian cMERA circuit (icMERA) has been proposed, whose higher order entangler operators  nonperturbatively generate genuine non-Gaussian trial states that go beyond the Gaussian ansatz of standard cMERA \cite{Haegeman:2011uy}. The icMERA evolution operator in the interaction picture for the scale parameter $u\in (u_{\text{IR}},u_{\text{UV}}]=(-\infty,0]$ reads
\begin{align}
\label{eq:icMERA_circuit}
U(u_1,u_2)
=
e^{-iu_1 L} \ {\cal P} e^{-i \int_{u_2}^{u_1} \hat K(u) du}\ e^{i u_2 L}
\ ,
\qquad
\hat K(u) = \hat K_0(u) +  \hat B(u) 
\ ,
\end{align}
where $L$ is a dilatation operator 
\begin{align}
L=
-\frac12 \int d\vx\left[
	\pi(\vx)(\vx\cdot\nabla\phi(\vx))
	+(\vx\cdot\nabla\phi(\vx))\pi(\vx)
	+\frac d2 (
		\phi(\vx)\pi(\vx)
		+\pi(\vx)\phi(\vx)
		)
	\right]
\ ,
\end{align}
and hatted operators $\hat\cO(u)$ denote operators $\cO(u)$ in the interaction picture,
\begin{align}
\hat \cO(u)
=
e^{iuL} \cO(u) e^{-iuL}
\ .
\end{align}
In particular, $\hat K_0(u)$ amounts to the Gaussian entangler of a free scalar theory given by the quadratic operator,
\begin{align}\label{eq:g-disentangler}
\hat K_0(u)
=
\frac12 \int d^dk  \ g(ke^{-u};u) \left[
	\phi(\vk)\pi(-\vk)
	+\pi(\vk)\phi(-\vk)
	\right]
\ ,
\end{align}
and $g(k;u)=g(u) \cdot \Gamma(k/\Lambda)$, where $g(u)$ is the variational parameter of the entangler. 
\footnote{Despite formally the variational parameter is $g(u)$, upon the parameterization of $g(k;u)$, the optimization of $g(k;u)$ straightforwardly reduces to that of $g(u)$. Throughout the text, we will unambiguously refer to both quantities as variational parameters.}
The function $\Gamma\left(k/\Lambda \right)$ is the typical cMERA hard cutoff $\Gamma(x)\equiv\Theta(1-x)$ with $\Theta$ the Heaviside step function.

In addition, the non-Gaussian entangler $\hat B(u)$,
\begin{align}\label{eq:icMERA_dis_in}
\hat B(u)
=
-s \int_{\vp\vq_1\cdots\vq_n} g(pe^{-u},q_1 e^{-u},\dots,q_ne ^{-u};u) \pi(\vp)\phi(\vq_1)\cdots\phi(\vq_n)\delta(\vp+\tsum_i\vq_i)
\ ,
\end{align}
$\mathbb{N}\ni n\ge2$,
nonperturbatively incorporates into the standard cMERA evolution non quadratic  terms through the variational function $g(p,q_1,\ldots,q_n;u)$ and the variational parameter $s$. Having a real wavefunctional requires time reversal invariance of the icMERA evolution which in turn amounts to having an odd number of $\pi$ operators in  $\hat B(u)$. Furthermore, as explained in \cite{Fernandez-Melgarejo:2020fzw}, $s$ is a variational parameter related to the coupling strength of the theory,  so the standard Gaussian cMERA evolution is recovered when $s \to 0$. Thus, together with the variational function $g(k;u)$ appearing in \eqref{eq:g-disentangler}, the new variational terms must be determined through energy minimization. 

Importantly, the variational functions $g(p,q_1,\ldots,q_n;u)$ and more concretely 
\begin{align}
f(p,q_1,\ldots,q_n;u)\equiv\int^u_0 d\sigma\, g(p e^{-\sigma},q_1 e^{-\sigma},\ldots,q_n e^{-\sigma};\sigma) \ ,
\label{eq:int-fs}
\end{align}
must fulfill a set of orthogonality constraints in order for icMERA to consistently work on the field operators.  This forces us to make some assumptions on the structure of these functions. While we refer the reader to the works \cite{Fernandez-Melgarejo:2019sjo,Fernandez-Melgarejo:2020fzw} for full details, here we briefly comment on this.

In \cite{Fernandez-Melgarejo:2020fzw} we proposed,
\footnote{Let us note that this is not the most general solution and other alternatives can also be considered, in combination with other not-so-sharpened cutoff functions.}
\begin{align}
g(p,q_1,\ldots,q_n;\sigma)
=
g_B(\sigma)\Gamma_B(p e^\sigma,q_1 e^\sigma,\ldots,q_n e^\sigma)\Gamma\left(\frac{p}{\Lambda }\right)\Gamma\left(\frac{q_1}{\Lambda }\right)\cdots \Gamma\left(\frac{q_n}{\Lambda }\right)
\ ,
\label{eq:g-ansatz-phi2}
\end{align}
where $g_B(\sigma)$ is the variational parameter that tunes the strength of the scale dependent non-Gaussian transformation and $\Gamma_B$ is a combination of cutoff functions depending on some variational cutoffs $\Delta_i$,
\begin{align}
\Gamma_B(p,q_1,\ldots,q_n)
=
\Gamma\left(\frac{p}{\Delta_1}\right)\left[
\Gamma\left(\frac{\Delta_1 }{q_1}\right)
-\Gamma\left(\frac{\Delta_2 }{q_1}\right)
\right]
\cdots\left[
\Gamma\left(\frac{\Delta_1 }{q_n}\right)
-\Gamma\left(\frac{\Delta_2 }{q_n}\right)
\right]
\ .
\label{eq:cutoffs}
\end{align}
The optimal function $\Gamma_{B}(p,q_1,\dots,q_2)$ has to be self-consistently found by determining both cutoffs $\Delta_i$. Namely, different from the Gaussian set-up, this scheme illustrates how the strength of the interaction variationally determines the region in momentum space that will be relevant in the optimization procedure. This feature turns out to be essential for strongly-coupled systems, which exhibit some regimes at which the Gaussian quasi-particle picture is no longer valid. 

The choice of \eqref{eq:g-ansatz-phi2} makes $f$ to fulfill the following orthogonality constraints, $\forall i=1,\ldots,n$,
\begin{align}
\label{eq:constraint}
f(q_i,q_1,\dots,q_n;\sigma)
=0 
\ , 
\qquad
f(p,q_1,\dots,q_n;\sigma)\, f(q_i,k_1,\ldots,k_n;\sigma) 
=&\ 0
\ ,
\end{align}
and thus obtain a consistent truncation for the action of icMERA on the field operators.

For a(n interacting) quantum field theory, let us consider the ground state of the theory to be $\ket{\Psi_\Lambda}\equiv\ket{\Psi(u=u_{\text{UV}})}\equiv\ket{\Psi_{\text{UV}}}$. As in standard cMERA, we define the state $\ket{\Omega}$ as that with no entanglement between spatial regions. $\ket\Omega\equiv\ket{\Psi(u=u_{\text{IR}})}$ is invariant with respect to spatial dilatations, which implies that $L\ket\Omega=0$. For an interacting theory, $\ket\Omega$ is such that
\begin{align}
\left(
	\sqrt M(\phi(\vk)-\chi_0)+\frac{i}{\sqrt M}\pi(\vk)
	\right) \ket\Omega
	= 0 
\ ,
\qquad
\quad
\chi_0=\expval{\phi}_\Omega
\ ,
\label{eq:Omega}
\end{align}
where $M=\sqrt{\Lambda^2+\mu^2}$ with $\mu$ a variational mass to be determined by the energy functional minimization. For the $\phi^4$ theory, $\mu$ satisfies the following gap equation:
\begin{align}
\mu^2
=
m^2
+\frac{\lambda}{2}\left(
	\int_\vk \left(k^2+\mu^2\right)^{-1/2}
	+\chi_0^2
	\right)
\ ,
\label{eq:mass-mu}
\end{align}
and $m$ and $\lambda$ are the bare couplings of the theory \cite{Cotler:2016dha}.

\subsubsection*{Non-Gaussian icMERA States}

\begin{figure}[!t]
\centering
\includegraphics[width=\textwidth]{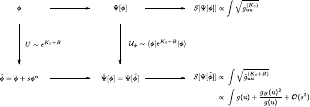}
\caption{
\textit{
Schematic relation among Gaussian and non-Gaussian objects. In the first line, upon assuming a field $\phi$ together with a Gaussian wavefunctional $\Psi[\phi]$, the entanglement entropy associated to this state is given by $S[\Psi[\phi]]$. In the second line, assuming the transformed field $\tilde\phi$, the non-Gaussian wavefunctional $\tilde\Psi[\phi]$ can be written as a Gaussian state in the $\tilde\phi$-basis. Consequently, the entanglement entropy is given by $S[\Psi[\tilde\phi]]$.
}}
\label{fig:scheme1}
\end{figure}

Let us consider $\ket\phi$ an eigenstate of the operator $\hat\phi$ with eigenvalue $\phi(\vk)$, 
\begin{align}
\hat\phi(\vk)\ket\phi=\phi(\vk)\ket\phi
\ .
\end{align}
Then the (coordinate) wavefunctional representation $\Psi[\phi]$ of a state $\ket\Psi$ is defined as
\begin{align}
\Psi[\phi] \equiv \braket{\phi}{\Psi}
\ .
\end{align}

If we consider a Gaussian state, $\ket{\Psi_G}$, then
\begin{align}
\Psi_G[\phi]
=
\braket{\phi}{\Psi_G}
= 
{\rm N}\exp\left(
		-\frac12 \int_{\vk} (\phi(\vk)-\chi_0)G^{-1}(\vk)(\phi(-\vk)-\chi_0)
	\right)
\ ,	
\label{eq:G-functional}
\end{align}
where $G^{-1}(\vk)=(\vk^2+m^2)$, ${\rm N}$ is a normalization constant.

Let us consider now the non-Gaussian wavefunctional $\tilde \Psi[\phi;u]$ generated by the transformation $U(0,u)$ acting on the aforementioned Gaussian state $\ket\Omega$,
\begin{align}
\tilde \Psi[\phi;u]
\equiv
\braket{\phi}{\tilde \Psi}=
\matrixel{\phi}{U(0,u)}{\Omega}
=
\matrixel{\phi}{P e^{-i \int_{u_{IR}}^{u} \hat K(\sigma) d \sigma}}{\Omega}
\ .
\end{align}
Consequently, when $U$ acts on $\ket\phi$, we have
\begin{align}
\tilde\Psi[\phi;u]
=
\matrixel{\phi}{U}{\Omega}
=
\int \cD\phi'\matrixel{\phi}{U}{\phi'}\braket{\phi'}{\Omega}
=
\cU_\phi(0,u)\Psi_{\Omega}[\phi]
\ ,
\end{align}
where $\Psi_{\Omega}[\phi]$ is the wavefunctional representation of $\ket\Omega$ and $\cU_\phi$ is the functional representation of the operator $U(0,u)$.
Interestingly, and this is the gist where the entanglement entropy calculation resides, the non-Gaussian variational functional $\tilde \Psi[\phi;u]$ is given by  \cite{Polley:1989wf,Ritschel:1990zs,Fernandez-Melgarejo:2020fzw}
\begin{align}
\tilde \Psi[\phi;u]
=
\Psi_{\Omega}[\tilde\phi;u]
\ ,
\end{align}
which is a Gaussian state in the basis $\tilde\phi(\vk)$, 
\begin{align}
\label{eq:tildephi-scalar}
\tilde\phi(\vk)
= 	
e^{-\frac{d}{2}u}
e^{-f(\vk;u)}\phi(\vk e^{-u})
	+s e^{n\frac{d}{2}u}\int_{\vq_1\cdots\vq_n} h(k,q_1,\ldots,q_n;u)\phi(\vq_1 e^{-u})\cdots\phi(\vq_n e^{-u})
\ ,
\end{align}
with
\footnote{At weak coupling, these results are qualitatively comparable with the ones obtained in perturbative cMERA approaches (\emph{cf.} (126), (149) in \cite{Cotler:2018ehb}).}
\begin{align}
h(k,q_1,\ldots,q_n;u)
\equiv
f(k,q_1,\ldots,q_n;u)\frac{e^{-f(k;u)}-\prod_{i=1}^n e^{-f(q_i;u)}}{f(k;u)-\sum_{i=1}^nf(q_i;u)}
\delta(p+\tsum_{i=1}^n q_i)
\ .
\label{eq:h-funct}
\end{align}
That is to say, $\Psi_{\Omega}[\tilde\phi;u]$ is a Gaussian state with respect to a nonlinear deformation of the infinite-dimensional configuration basis in such a way that $\Psi_{\Omega}[\tilde\phi;u]=\Psi_{\Omega}[\phi-s \phi^n;u]$. 

Interestingly, the fact that our non-Gaussian trial state can be written in terms of a Gaussian wavefunctional (expressed in another basis) enables us to use the standard formulas of the entanglement entropy for free theories. Such expressions, which apply to Gaussian states, have been written as a function of the parameters of a Gaussian cMERA circuit in \cite{Fernandez-Melgarejo:2021ymz}. Similarly, in the next section we will express the entanglement entropy in terms of the variational parameters of the non-Gaussian icMERA circuit.

\subsection{Entanglement Entropy}
\label{sec:scalar-ee}

For a Gaussian state described by the wavefunctional $\Psi[\phi]=\braket{\phi}{\Psi}$, the half space entanglement entropy is given by \cite{Solodukhin:2011gn}
\begin{align}
S_A\equiv S_A^{\Psi[\phi]}
=
\frac{|A_\perp|}{6}\int_\vp \log\expval{\phi(\vp)\phi(-\vp)}_{\Psi[\phi]}
\ ,
\label{eq:ee-Gaussian}
\end{align}
where $\expval{\cO}_{\Psi[\phi]}$ denotes the expectation value of the operator $\cO$ on the wavefunctional $\Psi[\phi]$ and $|A_\perp|$ is the area of the entangling surface for the half space.

On the other hand, when acting $U(0,u)$ on the field operator $\phi(\vk)$, we obtain a transformed field $\tilde\phi(\vk;u)$, 
\begin{align}
\phi(\vk) \to \tilde \phi(\vk;u) \equiv U^\dagger(0,u)\  \phi(\vk)\ U(0,u)
\ .
\end{align}
with $U(0,u)$ given in \eqref{eq:icMERA_circuit}. As a consequence, the wavefunctional associated to the transformed field $\tilde\phi$ can be written as
\begin{align}
\Psi[\phi] \to \tilde \Psi [\phi;u] \equiv \Psi[\tilde\phi;u]
\ ,
\label{eq:wavefunctionals}
\end{align}
where $\Psi[\tilde\phi;u]\equiv\Psi[\phi-s\phi^n;u]$ is a Gaussian wavefunctional with a nonlinearly displaced argument.

This implies that the entanglement entropy for a non-Gaussian wavefunctional $\tilde\Psi[\phi]=\Psi[\tilde\phi]=\braket{\tilde\phi}{\Psi}$ has the same expression as for a Gaussian state (\emph{cf.} \eqref{eq:ee-Gaussian}), but just written in terms of the transformed field $\tilde\phi(\vp)$,
\begin{align}
S_A^{\tilde\Psi[\phi]}
=
S_A^{\Psi[\tilde\phi]}
=
\frac{|A_\perp|}{6}\int_\vp \log\expval{\tilde\phi(\vp)\tilde\phi(-\vp)}_{\Psi[\tilde\phi]}
\ .
\end{align}
Because $\Psi[\tilde\phi;u]$ is actually a genuine Gaussian state in the $\tilde\phi$-basis, there exists a Gaussian cMERA circuit with an entangler $\hat{\tilde {K}}_0(u)$,
\begin{align}
\hat{\tilde {K}}_0(u)
=
\int_\vp \tilde g(k e^{-u};u) \left( \tilde\phi(\vp)\tilde\pi(-\vp) +\tilde\pi(\vp)\tilde\phi(-\vp)\right)
\ ,
\qquad
\tilde g(k;u)\equiv \tilde g(u) \Gamma(k/\Lambda)
\ ,
\label{eq:effective-K0}
\end{align}
where tilded operators are defined as $\tilde \cO\equiv U^\dagger \cO U$, such that it generates the exact ground state of the free theory of the field $\tilde\phi(\vk)$. According to the result of \cite{Fernandez-Melgarejo:2021ymz}, the entanglement entropy for the half space is entirely expressed in terms of the Gaussian cMERA variational parameter as
\begin{align}
\label{eq:ee-tilde}
\begin{split}
\tilde S_A\equiv S_A^{\Psi[\tilde\phi]}&= \cC_{d-1} \frac{|A_\perp|}{\epsilon^{d-1}}\, \int^0_{u_{\rm IR}}\, du\, \tilde g(u)\,  e^{u(d-1)} +\widetilde {\rm const}'
\\ &=\cC_{d-1} \frac{|A_\perp|}{\epsilon^{d-1}}\, \int^0_{u_{\rm IR}}\, du\, \sqrt{\tilde g_{uu}}\,  e^{u(d-1)} +\widetilde {\rm const}'
\, .
\end{split}
\end{align}
where $\cC_{d-1} =\frac13\frac{S^{(d-1)}}{(d-1)}$, $S^{(d)} \equiv \frac{2\pi^{d/2}}{\Gamma(d/2)}$  is the area of a $d$-sphere, $\epsilon \equiv 1/\Lambda$, with $\Lambda$, the momentum cutoff in cMERA and  $\tilde g_{uu}(u)$ is the Fisher information metric given by 
\begin{align}
\tilde g_{uu}(u) du^2
=
\cN^{-1}\left(
	1
	-\left|
		\braket{\tilde\Psi(u)}{\tilde \Psi(u+du)}
		\right|^2
	\right)
\ ,
\quad
\ket{\tilde\Psi(u)}
=
P e^{-i \int_{u_{IR}}^{u} \hat{\tilde{ K}}_0(\sigma) d \sigma}\ket{\Omega}
\ ,
\label{eq:fisher}
\end{align}
and $\cN$ is a normalization constant $\cN=\delta^d(0)\, (\Lambda e^u)^{(d-1)}$, where $\delta^d(0)$ is the (infinite) volume of the $d$-dimensional space $\bbR^d$. On the other hand, $\ket{\tilde\Psi(u)}$ is written in terms of an icMERA circuit as follows:
\begin{align}
\ket{\tilde\Psi(u)}
=
P e^{-i \int_{u_{IR}}^{u} \hat{ K}(\sigma) d \sigma}\ket{\Omega}
\ ,
\qquad
\hat K(u)
\equiv
\hat K_0(u)
+
\hat B(u)
\ ,
\end{align}
for which the Fisher information metric $\tilde g_{uu}(u)$ is
\begin{align}
\tilde g_{uu}
=
\cN^{-1}\, \left(\expval{\hat K(u)^2}_\Omega 
	-\expval{\hat K(u)}^2_\Omega\right)
\ .
\end{align}

The result \eqref{eq:ee-tilde} has been obtained by solely considering the Gaussian nature of the icMERA wavefunctional in the $\tilde \phi$-basis and the results for the half space entropy in a Gaussian cMERA \cite{Fernandez-Melgarejo:2021ymz}. The entanglement entropy for Gaussian states in half space entirely depends on the 2-point correlator. Such correlator is replaced by the Fisher metric obtained from the icMERA circuit instead of the one obtained from the Gaussian entangler $\hat{\tilde{K}}_0$. Noteworthily, Eq. \eqref{eq:ee-tilde} suggests that $\tilde{g}(u)$ consists of a local density of disentanglers (\emph{a.k.a}. bond dimension) of the icMERA circuit. 

Thus, in order to generalize our previous results for the case of a Gaussian cMERA, we will show that this local bond dimension reduces to the Fisher metric of the circuit. 
That is to say, we will firstly calculate the Fisher metric for various icMERA entanglers at strong coupling. By plugging these results in \eqref{eq:ee-tilde}, we will obtain the entropy of the half space in different interaction regimes. 
In Section \ref{sec:ON-ee} we will apply the same method for theories with $N$ scalar fields and $O(N)$ symmetry. 

\subsection{Bond Dimension in cMERA}
\label{sec:scalar-fidelity}

In a discrete tensor network, the indices connecting the tensors represent the structure of the many-body entanglement in a quantum state. The number of values that these indices can take are in correspondence with the correlation strength that is encoded in such state. The maximum number of values that these indices take is called the bond dimension $J$ of the tensor network. 

For a tensor network state and a bipartition of the system, the entanglement entropy is then upper bounded by \cite{Nozaki:2012zj}
\begin{align}
S_A
\le \#\text{bonds}(\gamma_A)\, \log J
\ ,
\end{align}
where $\gamma_A$ is the curve that divides the system and $\#\text{bonds}(\gamma_A)$ is the number of bonds intersected by the entangling curve $\gamma_A$. For the case of the half space, the entanglement entropy in the discrete MERA tensor network representing a $(d+1)$-dimensional interacting quantum system on a lattice reads  \cite{Nozaki:2012zj}
\begin{align}
  S_A\propto L^{d-1}\sum_{u=-\infty}^0 n(u)\cdot 2^{(d-1)u}\, ,
  \label{eq:ee_dmera}
\end{align}
where $L^{d-1}$ is the number of lattice points on the boundary of $A$ and $n(u)$
measures the strength of the bonds at the layer specified by the non-positive integer $u$. In other words, $n(u)$ amounts to the logarithm of the bond dimension at the layer $u$. Thus, it is straightforward to see that expression \eqref{eq:ee-tilde} provides a continuum generalization of \eqref{eq:ee_dmera} for Gaussian states, where the half space entanglement entropy can be naturally written as the integral of the cMERA variational parameter $g(u)$, the density of disentanglers, along the scale parameter. Given that \eqref{eq:ee_dmera} is valid for non-Gaussian states and our previous discussion on the icMERA states, we will interpret the Fisher metric of an (i)cMERA state as the local \emph{bond dimension} for the circuit at a given scale $u$.

In this section, we are going to review the bond dimension in Gaussian cMERA states and calculate the bond dimension of the icMERA circuit for various entanglers. This quantity, which is a function of the renormalization scale $u$ through the variational parameters $g(u)$ and $g_B(u)$, must be determined upon the energy minimization of the specific interacting theory. When integrated along the renormalization scale, we will obtain the entanglement entropy associated to the (non-)Gaussian state living in the UV limit of the tensor network.

\subsubsection{Bond dimension in Gaussian cMERA}

The entanglement entropy for the half space in Gaussian cMERA has recently been obtained in a very natural fashion as the integral of the density of the entanglers along the curve that partitions the system \cite{Fernandez-Melgarejo:2021ymz}. The local density of entanglers, which, as commented above, is an avatar of the bond dimension in the discrete versions of MERA, is given by $\sqrt{g_{uu}}$, where $g_{uu}$ is the Fisher metric \cite{Nozaki:2012zj},
\begin{align}
g_{uu}(u) du^2
=
\cN^{-1}\left(
	1
	-\left|
		\braket{\Psi(u)}{\Psi(u+du)}
		\right|^2
	\right)
\ ,
\label{eq:fisher2}
\end{align}
with
\begin{align}
\ket{\Psi(u)}
= 
P e^{-i \int_{u_{IR}}^{u} \hat K_0(\sigma) d \sigma}\ket{\Omega}
\ ,
\label{eq:Phi_G}
\end{align}
with $\hat K_0(\sigma)$ being the Gaussian \emph{entangler} operator in the interaction picture defined in \eqref{eq:g-disentangler}. As stated above, $g_{uu}$ results:
\begin{align}
g_{uu}(u)
=
\cN^{-1}\, \left(\expval{\hat K_0(u)^2}_\Omega 
	-\expval{\hat K_0(u)}^2_\Omega\right)= g(u)^2
\ .
\label{eq:1-fidelity}
\end{align}

\subsubsection{Bond dimension in non-Gaussian icMERA}
\label{sec:fidelity_ex}

The above result can be generalized to the case in which the (non-)Gaussian cMERA circuit is made up of more entangling operators,
\begin{align}
\hat K(u)=\sum_i\, \hat K_i(u)\, , \qquad \hat K_i(u)=\int_{\vk}\, g_i(u)\cdot\Gamma_{i}(k_j;u)\, \hat\cO_i(\vk_j)\
\ ,
\label{eq:sum-entanglers}
\end{align}
with $\hat K_i$ an entangler defined in terms of the operator $\hat \cO_i$ and a generic function $g_i(k_j;u)$ containing the variational parameter $g_i(u)$, both of them depending on a set of momenta $\vk_j$. In this case, the Fisher metric $g_{uu}$ is a quadratic function of the variational parameters,
\begin{align}
g_{uu} \propto \sum_{ij}  \langle \hat K_i(u)\, \hat K_j(u)\rangle_{\Omega}\propto \sum_{ij} g_i(u) g_j(u)
\ .
\label{eq:sum-metric}
\end{align}
The structure \eqref{eq:sum-metric} shows that, given the r\^ole of the Fisher metric as a local  bond dimension in cMERA,  adding more entanglers, and consequently having more terms in the sum, is the analog of increasing the bond dimension in the discrete MERA circuit. In this respect, the icMERA tensor network discussed above  provides a set of nontrivial examples for the increasing  of the bond dimension in cMERA--type circuits by naturally incorporating additional non quadratic entanglers due to the non-Gaussian trial states generated to tackle interacting theories.

Let us consider an icMERA circuit given by the disentangler,
\begin{align}
{\hat K}(u)={\hat K_0}(u) + {\hat B}(u)
\ ,
\end{align}
with ${\hat K_0}(u)$ given in \eqref{eq:g-disentangler} and ${\hat B}(u)$ given by a generic disentangler of the form \eqref{eq:icMERA_dis_in} for which we will use the acronym $\hat B = \pi \phi^n$.\footnote{For concrete examples in obtaining optimal values for the variational parameters of the icMERA circuit, we refer to \cite{Fernandez-Melgarejo:2019sjo, Fernandez-Melgarejo:2020fzw} where extensive treatments on the $\lambda \phi^4$ theory with a $\hat B = \pi \phi^2$ disentangler have been carried out.}

Now, we will calculate the Fisher information metric for the state
\begin{align}
\ket{\Psi(u)}
= 
P e^{-i \int_{u_{IR}}^{u} \hat K(\sigma) d \sigma}\ket{\Omega}
\ ,
\label{eq:Phi_NG}
\end{align}
Applying the formula \eqref{eq:1-fidelity} to $\hat K(u)$, when calculating $\expval{\hat K(u)}_\Omega$ we obtain:
\footnote{While it is straightforward to prove that $\expval{\hat K_0(u)}_\Omega=0$, proving that $\expval{\hat B(u)}_\Omega$ vanishes requires the application of Wick's theorem. Consequently,  we always find $\delta$-functions of the type $\delta(\vk\pm \vq_i)$, where $\vk$ and $\vq_i$ are the momenta carried by $\pi$ and $\phi$, respectively. Then, because of the orthogonality conditions of $f(p,q_1,\cdots,q_n)$, every term vanishes.}
\begin{align}
\expval{\hat K(u)}_\Omega
=
\expval{\hat K_0(u)}_\Omega
+\expval{\hat B(u)}_\Omega
=
0 
\ .
\end{align}
Regarding $\expval{\hat K(u)^2}_\Omega$ we have
\begin{align}
\expval{\hat K(u)^2}_\Omega
=\expval{\hat K_0(u)^2}_\Omega
+\expval{\hat K_0(u)\hat B(u)}_\Omega
+\expval{\hat B(u)\hat K_0(u)}_\Omega
+\expval{\hat B(u)^2}_\Omega
\ .
\label{eq:fidelity-R2}
\end{align}

Thanks to the calculability bonanza of the icMERA ansatz, it is possible to compute each term separately using Wick's theorem. The first one was known from the free case and reads 
\begin{align}
\expval{\hat K_0(u)^2}_\Omega
=&\
\cN\, g(u)^2
\ .
\end{align}
The rest of the terms will be calculated in the next paragraphs for different entanglers.

\subsubsection*{Entangler $\hat B\equiv \pi\phi^2$}

Assuming $n=2$ in the expression \eqref{eq:icMERA_dis_in} for $\hat B(u)\sim \pi\phi^2$, we have
\begin{align}
\hat B(u)
=
-s \int_{\vp\vq_1\vq_2} g(pe^{-u},q_1 e^{-u},q_2e ^{-u};u) \pi(\vp)\phi(\vq_1)\phi(\vq_2)\delta(\vp+\vq_1+\vq_2)
\ .
\label{eq:B-piphi2}
\end{align}
This operator induces additional terms to those obtained from a Gaussian state when evaluating the expectation value of the Hamiltonian. That is to say, if the original theory  had a $\mathbb{Z}_2$ symmetry, it is broken by the new terms generated by $\hat B(u)$ \cite{Fernandez-Melgarejo:2020fzw}. Thus, in case the ground state is in a symmetric phase, the icMERA circuit uses the degrees of freedom of the symmetry-broken phase in order to minimize the energy.  This can be easily understood in the weak coupling regime of the $\phi^4$ theory, where these new  terms in the Hamiltonian vev correspond to (variationally optimized) vertices in Feynman diagrams that give account of non-Gaussian corrections to the connected part of the 2-point and 4-point correlation functions \cite{Fernandez-Melgarejo:2020fzw}. 

Then, evaulating \eqref{eq:fidelity-R2} we obtain that, while the crossed terms vanish
\begin{align}
\expval{\hat K_0(u)\hat B(u)}_\Omega
=
\expval{\hat B(u)\hat K_0(u)}_\Omega
=
0
\ ,
\end{align}
the quadratic contribution results
\begin{align}
\expval{\hat B(u)^2}_\Omega
&=
\delta^{d}(0)\, \frac{s^2}{8 M}\, \left[\phi_c^2 + 2\, \bar \chi_2(u)\right]
\, ,
\label{eq:B2-phi2}
\end{align}
where  $M=\sqrt{\Lambda^2+\mu^2}$, with $\mu$ the variational mass defined in \eqref{eq:mass-mu}. In addition, we have defined
\begin{align}
\phi_c=\expval{\tilde\phi}_{\Omega} \equiv \bar \chi_1(u) 
= 
\int_{\vq}c\left(\vq e^{-u},-\vq e^{-u};u\right)
\, ,
\qquad
c(\vp,\vq;u)\equiv g(|\vp + \vq|,p, q;u)
\, .
\end{align}
As $\bar \chi_1$ depends on several parameters of the ansatz, defining the corresponding combination as a new parameter $\phi_c$, one of the original variables can be eliminated. This new variable is very convenient in order to solve the optimization equations of the circuit. 
They can be easily solved for $\phi_c=\bar\chi_1(u)\sim 0$, which amounts to the disconnected contribution of the 2-point correlators \cite{Ritschel:1990zs,Fernandez-Melgarejo:2020fzw}.

On the other hand, $\bar\chi_2$ is a loop integral given by\footnote{A bar $\bar\chi$ has been added to these momentum integrals in order to differentiate them from the unbarred momentum integrals $\chi$'s entering the position space correlators in \cite{Fernandez-Melgarejo:2019sjo,Fernandez-Melgarejo:2020fzw}.}
\begin{align}
\bar \chi_2(u) =  \int_{\vp,\vq}c\left(\vp e^{-u},\vq e^{-u};u\right)^2\, .
\label{eq:chi1chi2}
\end{align}
$\bar\chi_2(u)$ describes its connected part, as it can be seen from its relation to the momentum integral $\chi_2(\vp;u)$ in \cite{Fernandez-Melgarejo:2020fzw}.

Therefore, the Fisher metric of the $\pi\, \phi^2$ icMERA circuit, $\tilde g_{uu}$, is then given by
\begin{align}\label{eq:fisher_icMERA1}
\tilde{g}_{uu}\equiv  \tilde{g}(u)^2= g(u)^2 + \gamma^2\, e^{-u(d-1)}\left[\phi_c^2 + 2\, \bar \chi_2(u)\right]\, ,
\qquad
\gamma^2 \equiv \frac{s^2}{8 M\, \Lambda^{(d-1)}}
\ ,
\end{align}
whose expression exhibits the aforementioned structure \eqref{eq:sum-metric}. In addition, the Fisher metric from the Gaussian cMERA is trivially recovered for $s=0$.

From this formula and, according to the results obtained in Section \ref{sec:scalar-ee}, the bond dimension for an \emph{effective} Gaussian field reads
\begin{align}
\tilde g(u) = \sqrt{g(u)^2 + \gamma^2\, e^{-u(d-1)}\, \left[\phi_c^2 + 2\, \bar \chi_2(u)\right]}
\, .
\end{align} 
Given our discussions above, this expression can be straightforwardly plugged in the entanglement entropy formula \eqref{eq:ee-tilde} to obtain the half space entanglement entropy of a self interacting scalar theory whose ground state is approximated by an icMERA circuit with $B = \pi\, \phi^2$.

For the case of the $\lambda\, \phi^4$ theory, as one solves the optimization equations for a fixed value of $\phi_c\equiv \expval{\phi}\sim 0$, the parameter $s$ results proportional to the coupling $s \sim \lambda\, \phi_c$ \cite{Fernandez-Melgarejo:2020fzw}. For the sake of convenience, we can consider the limit $\phi_c\sim 0$, which in turn implies that $\gamma\ll1$ even for a large but finite coupling $\lambda$. Namely, assuming $\gamma$ small does not necessarily imply that we are considering the theory at weak coupling. That said, in this limit $\tilde g(u)$ results
\begin{align}\label{eq:geff_piphi2}
\tilde g(u) 
= g(u) +\gamma^2  g_{\rm qu}(u) +\cO(\gamma^4)\, ,
\quad\qquad
g_{\rm qu}(u)
\equiv\, e^{-u(d-1)}\frac{\left[\phi_c^2 + 2\, \bar \chi_2(u)\right]}{2\, g(u)}
\ .
\end{align}
Two terms contribute to $ \tilde g(u)$. The former consists of the standard variational parameter which is sourced by the Gaussian entangler $\hat K_0(u)$ inside the icMERA circuit \cite{Haegeman:2011uy}. Consequently, it can be understood as a semiclassical contribution. In contrast, the second term is proportional to $\gamma^2$ and represents the quantum corrections to the \emph{bond dimension}. Namely, the non-Gaussian transformations induce new contributions in the correlators which, when taking the weak coupling limit, can be understood as arising from variational effective vertices \cite{Fernandez-Melgarejo:2020fzw}. 

In order to clarify the r\^ole of the variational parameters in  $\tilde g(u)$, the result \eqref{eq:geff_piphi2} can be tidied up by splitting apart the momentum integrals. Recalling that $c(\vp e^{-u},\vq e^{-u})$ can be written as
\begin{align}
c(\vp e^{-u},\vq e^{-u})=g_{\cB}(u)\cdot \Gamma_{B}(|\vp + \vq|, p, q)\cdot \Gamma\left(|\vp +\vq| e^{-u}/\Lambda\right)\, \Gamma\left(p e^{-u}/\Lambda\right)\, \Gamma\left(q e^{-u}/\Lambda\right)\, ,
\end{align}
we have
\begin{align}
\bar \chi_{2}(u)= g_{B}(u)^2 \cdot I_2(\Lambda e^{u})\, ,
 \end{align}
where $I_{2}$ is an integral over momenta of the $\Gamma$-cutoff functions:
\begin{equation}
\begin{split}
I_2(\Lambda e^{u}) =&\ \int_{\vp,\vq}\, \Gamma_{B}(|\vp + \vq|, p,q)\, \Gamma\left(|\vp +\vq| e^{-u}/\Lambda\right)\, \Gamma\left(p e^{-u}/\Lambda\right)\, \Gamma\left(q e^{-u}/\Lambda\right)\, .
\end{split}
\end{equation}
As a result, the half-space entanglement entropy in icMERA with $B = \pi\, \phi^2$ reads
\begin{align}
\label{eq:ee_piphi2}
\tilde S_A = S^{(0)}_A + s^2\, S^{\chi}_A + \cO(s^4)\, ,
\end{align}
where
\begin{align}
S_A^{(0)} =&\ \cC_{d-1} \frac{|A_\perp|}{\epsilon^{d-1}}\, \int^0_{u_{\rm IR}}\, du\,  g(u)\,  e^{u(d-1)} +\widetilde {\rm const}'\,  ,
\\
S^{\chi}_A =&\ \frac{\cC_{d-1}}{8\, M} \, |A_\perp|\, \int^0_{u_{\rm IR}}\, du\, \left(\frac{2\,g_{B}(u)^2\, I_2(\Lambda e^{u}) +\phi_c^2 }{2 g(u)}\right)
\, .
\end{align}

\subsubsection*{Entangler $\hat B \equiv \pi\, \phi^3$ ($\mathbb{Z}_2$ symmetric)}
Let us take now $n=3$ in \eqref{eq:icMERA_dis_in}, \emph{i.e.}, let us consider the non-Gaussian entangler $\hat B \sim \pi\, \phi^3$,
\begin{align}
\hat B(u)
=
-s \int_{\vp\vq_1\vq_2\vq_3} g(pe^{-u},q_1 e^{-u},q_2e ^{-u},q_3e ^{-u};u) \pi(\vp)\phi(\vq_1)\phi(\vq_2)\phi(\vq_3)\delta(\vp+\vq_1+\vq_2+\vq_3)
\ .
\label{eq:B-piphi3}
\end{align}
This entangler induces additional terms in the vacuum expectation value of the interacting Hamiltonian in such a way that, if a $\mathbb{Z}_2$ symmetry exists, it is preserved \cite{Fernandez-Melgarejo:2020fzw}.

The Fisher metric \eqref{eq:1-fidelity} for this icMERA circuit is evaluated to obtain
\begin{align}
\tilde g_{uu}(u)=g(u)^2 + \gamma^2\, e^{-u(d-1)}  \, \bar \zeta_1(u)\, ,
\qquad\quad
\gamma^2\equiv \frac{s^2}{M^2\, \Lambda^{(d-1)}}
\ ,
\label{eq:guu-scalar-phi3}
\end{align}
where $\bar\zeta_1(u)$, which is related to the non-Gaussian corrections of the 2-point functions \cite{Ritschel:1990zs}, is given by
\begin{align}
\bar \zeta_1(u)=\frac{3}{4}\, \int_{\vp \vq \vr}\, c(\vp e^{-u},\vq e^{-u},\vr e^{-u};u)^2
\ ,
\qquad
c(\vp,\vq,\vr;u)=g(|\vp+\vq+\vr|,p,q,r;u)
\ ,
\end{align}
as defined in \eqref{eq:g-ansatz-phi2}. When separating the momenta and the renormalization scale integrals, $\bar\zeta_1(u)$ can be rewritten as
\begin{align}
\bar \zeta_1(u)=\frac{3}{4}\, g_{B}(u)^2\cdot I_{3}(\Lambda e^{u})\, ,
\end{align}
where $I_3$ is an integral over momenta
\begin{align}
I_{3}(\Lambda e^{u}) = \int_{\vp \vq \vr}\, \Gamma_{B}(|\vp+\vq+\vr|,p,q,r)\, 
\Gamma\left(\frac{|\vp+\vq+\vr|e^{-u}}{\Lambda}\right)
\Gamma\left(\frac{p e^{-u}}{\Lambda}\right)\Gamma\left(\frac{q e^{-u}}{\Lambda}\right)\Gamma\left(\frac{r e^{-u}}{\Lambda}\right)
\,.
\end{align}
When substituting in \eqref{eq:guu-scalar-phi3}, 
\begin{align}
\tilde g_{uu}(u)=g(u)^2 + \gamma^2\, \frac{3}{4}\, \, e^{-u(d-1)}\, I_{3}(\Lambda e^{u}) g_B(u)^2
\ ,
\end{align}
the Fisher information metric consists of the sum of the squared variational parameters of the circuit, as it was predicted above \eqref{eq:sum-metric}.

Finally, in the limit $s\ll1$, the non-Gaussian icMERA bond dimension is written as
\begin{align}
\label{eq:geff_icmera_boxed_pf3}
\tilde g(u) = g(u) + \frac{3 s^2}{8 M^2\, \Lambda^{(d-1)}}\, 
\, e^{-u(d-1)}\left(\frac{g_{B}(u)^2}{g(u)}\right)\cdot I_{3}(\Lambda e^{u})
+
\cO(s^4)
\, ,
\end{align}
which, as a result, gives the half-space entanglement entropy that reads
\begin{align}
\label{eq:ee_piphi3}
\tilde S_A = S^{(0)}_A + s^2\, S^{\zeta}_A + \cO(s^4)\, ,
\end{align}
with
\begin{align}
S^{\zeta}_A &= \frac{3\, \cC_{d-1}}{8\, M^2} |A_\perp|\, \int^0_{u_{\rm IR}}\, du\, \left(\frac{g_{B}(u)^2}{g(u)}\right)\cdot I_{3}(\Lambda e^{u})\, .
\end{align}

Hence, in this section we have obtained the half space entanglement entropy from the Fisher information metric generated by the icMERA circuit. The result, for various entanglers, is in agreement with the interpretation of the Fisher metric as the bond dimension of a tensor network: it consists of the sum of the squares of the variational parameters entering the circuit. Namely, it provides the density of entanglers cutting bonds at each scale and its sum (integral) along the renormalization scale produces the entanglement entropy.

More specifically, our results show that, in the limit $s \ll 1$, for two choices of the non-Gaussian part of the circuit, the entanglement entropy converges up to subleading corrections, while the leading term $S^{(0)}_A$ associated to the Gaussian part of the circuit exhibits an area law kind of divergence \cite{Fernandez-Melgarejo:2021ymz}. 

In the next section we are going to formulate icMERA for scalar theories with $N$ fields and $O(N)$ symmetry to study the large $N$ limit behavior of both the bond dimension and the entanglement entropy.

\section{Scalar $O(N)$ Model}
\label{sec:ON-icmera}

In this section we will introduce the icMERA circuit, which consists of a renormalization method for theories with $N$ scalar fields and global $O(N)$ symmetry. Our goal is to calculate the half space entanglement entropy upon considering that the non-Gaussian trial states generated by this circuit are Gaussian states expressed in a different field basis. Finally, we will show that this entanglement entropy can also be given in terms of the Fisher information metric. The latter, which is interpreted as the bond dimension of the cMERA circuit for an interacting theory, connects with the MERA tensor network perspective in which the entropy is captured by summing over the bond dimensions of tensors lying on a bulk entangling surface.

Let us firstly consider some simple examples of a scalar theory with $O(N)$ symmetry.  Here the degrees of freedom are $N$ scalar fields transforming in the fundamental representation, $\bsPhi \equiv \vec \phi =\left( \phi_1,\, \phi_2\, \cdots \phi_N\right)$.  
The simplest example of these models is given by the free theory
\begin{align}
\mathcal{L} = \frac{1}{2}\left[(\partial\, \bsPhi)^2 - m^2 (\bsPhi)^2\right]\, ,
\end{align}
with correlators $\langle \phi_i \phi_j \rangle = \delta_{ij}$ and commutation relations 
\begin{align}
[\phi_i(\vp),\pi_j(\vq)]=i\delta_{ij}\delta(\vp+\vq)
\ ,
\quad
[\phi_i(\vp),\phi_j(\vq)]=[\pi_i(\vp),\pi_j(\vq)]=0
\ ,
\quad
\ 
i,j=1,\ldots,N
\ ,
\end{align}
so that the theory is equivalent to $N$ copies of a free scalar field. Interaction terms can be included via a scalar potential $V(\bsPhi\cdot\bsPhi)$, for instance, a quartic term as 
\begin{align}
\mathcal{L}  = \frac{1}{2}\left[(\partial\, \bsPhi)^2 - m^2 (\bsPhi)^2\right] - \lambda (\bsPhi \cdot \bsPhi)^2\, ,
\label{eq:ON_theory}
\end{align}
The quartic interaction term scales as $N^2$,  while both the kinetic and mass terms scale as $N$. To keep the action to order $N$ when $N\rightarrow \infty$, $\lambda$ must be substituted by a new coupling $\tilde \lambda$ such that $\tilde \lambda \equiv N \lambda$ is finite.  Under this choice, the full action is proportional to $N$ and the theory becomes solvable in a saddle point approximation.  

Despite in this paper we will exclusively work this interaction model, the icMERA circuits that will be introduced in the following section, can be applied to any scalar potential. In this case, the scaling in the coupling as to be modified in such a way that the same criterion is fulfilled. 

Along this section we will follow the same notation as in Section \ref{sec:scalar-ee} to denote the (non-) Gaussian entanglers.

\subsection{Gaussian $O(N)$ cMERA}

Let us firstly consider the Gaussian part of the cMERA disentangler which, in the interaction picture, reads
\begin{align}
\hat K^{(0)}(u)
&=
\frac12 \int_\vk g(k e^{-u};u)\left[ \bsPi(\vk)\cdot\bsPhi(-\vk)+\bsPhi(\vk)\cdot\bsPi(-\vk)\right]
\ .
\label{eq:entangler-ON-G}
\end{align}
This entangler, which preserves the $O(N)$ symmetry, can be applied to the Hamiltonian of both the free and interacting theory. For the latter, the interaction contributions will exclusively affect the variational mass $\mu$ through the gap equation, as it occurs in the scalar case \cite{Cotler:2016dha}.

Nevertheless, and for later convenience, in the construction of the interacting disentangler, let us consider the case in which one of the components of the $O(N)$ vector $\bsPhi$ has a nonvanishing vacuum expectation value. As explained in \cite{Stevenson:1986nb} the most general Gaussian wave functional will have different kernel functions for the ``radial'' ($R$) and ``transverse'' ($T$) fields, where the ``radial'' direction is defined as the direction picked out by the classical field. In this case, there is a remaining $O(N-1)$ symmetry and, for $\bsPhi=(\phi_R,\vphi_T)$,  the Gaussian entangler in the interaction picture as
\begin{multline}
\hat \cK^{(0)}(u)
=
\frac12 \int_\vk \ g_R(k e^{-u};u)\left(
	\phi_R(\vk)\pi_R(-\vk)+\pi_R(\vk)\phi_R(-\vk)
	\right)
\\
+\ g_T(k e^{-u};u)\left(
	\vec\phi_T(\vk)\cdot\vec\pi_T(-\vk)+\vec\pi_T(\vk)\cdot\vec\phi_T(-\vk)\right)
\ ,
\label{eq:entangler-ON-G-RT}
\end{multline}
where $\vec \phi_T, \vec \pi_T$, \ldots are
 $O(N - 1)$ vectors, $g_{R(T)}(k ;u)$ can be assumed to be of the form
\begin{align}
g_{R(T)}(k ;u) =
 g_{R(T)}(u) \cdot \Gamma(ke^{-u}/\Lambda)
\ ,
\end{align}
and $g_{R(T)}(u)$ is the Gaussian density of disentanglers (i.e., bond dimension) in the radial (transverse) directions.

The $O(N)$ Gaussian cMERA evolution operator in the interaction picture for the entanglers \eqref{eq:entangler-ON-G} and \eqref{eq:entangler-ON-G-RT} with scale parameter $u\in(u_{\text{IR}},u_{\text{UV}}]=(-\infty,0]$ reads, respectively,
\begin{align}
U^{(0)}(u_1,u_2)
=
e^{-iu_1 L} \ {\cal P} e^{-i \int_{u_2}^{u_1} \hat K^{(0)}(u)du}\ e^{i u_2 L}
\ ,
\quad
\cU^{(0)}(u_1,u_2)
=
e^{-iu_1 L} \ {\cal P} e^{-i \int_{u_2}^{u_1} \hat \cK^{(0)}(u)du}\ e^{i u_2 L}
\ .
\label{eq:ON-cMERA}
\end{align}

When applying, for example, $\cU^{(0)}(0,u)$ to the above interacting theory, Eq. \eqref{eq:ON_theory}, the solution of the variational parameters imposed by the energy minimization of the energy functional is given by \cite{Stevenson:1986nb}
 \begin{eqnarray}\label{eq:gN_ddis}
 g_{R}(k;u)&=&\f{1}{2}\cdot \frac{e^{2u}}{e^{2u}+\mu_R^2/\Lambda^2}\cdot \Gamma(k/\Lambda)\, ,\\ \nonumber
 g_{T}(k;u)&=&\f{1}{2}\cdot \frac{e^{2u}}{e^{2u}+\mu_T^2/\Lambda^2}\cdot \Gamma(k/\Lambda)\, ,
 \end{eqnarray}
with  the variational mass parameters
 \begin{eqnarray}
 \mu_R^2 &=&m^2 + \lambda \left((N+1)I^T_0 + I^R_0 + \phi^2_c\right)\, ,\\ \nonumber
 \mu_T^2 &=& m^2 + \lambda \left((N-1)I^T_0 + 3I^R_0 + 3\phi^2_c\right)\, ,
\end{eqnarray}
where $m,\, \lambda$ are the bare mass and coupling constants of the theory respectively, $\phi_c \equiv\langle \Psi_{\Lambda}|\phi_R|\Psi_{\Lambda}\rangle$ and $I^{R,T}_0=1/2\,\int_{\vk}\, \left(\vk^2 +\mu_{R,T}^2\right)^{-1/2}$. As expected, when $\phi_c \sim 0$, we find $\mu_R = \mu_T$ and the $O(N)$ symmetry is restored.
 
As a further step, let us consider now some entanglers that generate non-Gaussian trial states to tackle interacting theories. In this case, we consider two types of entanglers, depending on whether $O(N)$ is preserved or not.

\subsection{$O(N)$ icMERA Circuit}

Let us introduce the $O(N)$ symmetric operator $\hat B_N(u)$
\begin{align}
B_N(u)
&=
\beta \int_{\vq_i} g(q_1,q_2,q_3,q_4;u)\, \bsPi(\vq_1)\cdot\bsPhi(\vq_2) \ \bsPhi(\vq_3)\cdot\bsPhi(\vq_4)\, \delta(\tsum_i\vq_i)
\ .
\label{eq:entangler-ON-G-sym}
\end{align}
In this case, the scale dependent function $g(q_1,q_2,q_3,q_4;u)$ contains the variational parameter $g_B(u)$:
\begin{align}
g(q_1,q_2,q_3,q_4;u)
=
g_B(u)\cdot \Gamma(q_1/\Lambda)\cdots \Gamma(q_4/\Lambda) \Gamma_B(\vq_1,\vq_2,\vq_3,\vq_4)
\ ,
\end{align}
where $\Gamma_B(\vq_1,\vq_2,\vq_3,\vq_4)$ is given in \eqref{eq:cutoffs}. This entangler can be understood as the straightforward generalization of the entangler \eqref{eq:icMERA_dis_in} for the case of $N$ fields when $n=3$.

However, we can avoid the presence of $\Gamma_B$ by considering that $O(N)$ is spontaneously broken down to $O(N-1)$ and then using the following entangler:
\begin{align}
\hat \cB_N(u)
=
\beta \int_{\vq_i} g(q_1e^{-u},q_2e^{-u},q_3e^{-u};u)\ \pi_R(\vq_1) \ \vec \phi_T(\vq_2)\cdot\vec \phi_T(\vq_3) \delta(\tsum_i\vq_i)
\ ,
\label{eq:entangler-ON-NG}
\end{align}
where the variational parameter $g_\cB(u)$ is contained in
\begin{align}
g(q_1,q_2,q_3;u)
=
g_\cB(u)\cdot \Gamma(q_1/\Lambda)\Gamma(q_2/\Lambda) \Gamma(q_3/\Lambda)
\ .
\end{align}
In addition, $\beta$ is a variational parameter which, as it will be shown, is proportional to the coupling constant. Preserving some orthogonal symmetry requires expressing the entangler in terms of inner products of vector fields, which in turn implies having an even power of scalar fields. We will elaborate on this and its effects on the $1/N$ expansion in the following sections.

Therefore, for a scale parameter $u\in(u_{\text{IR}},u_{\text{UV}}]=(-\infty,0]$, the unitary non-Gaussian $O(N)$ icMERA evolution operator is built upon the Gaussian, (\eqref{eq:entangler-ON-G} or \eqref{eq:entangler-ON-G-RT}) and the non-Gaussian entanglers (\eqref{eq:entangler-ON-NG} or \eqref{eq:entangler-ON-NG}) as follows:
\begin{align}
U(u_1,u_2)
=&\
e^{-iu_1 L} \ {\cal P} e^{-i \int_{u_2}^{u_1} \hat K(u) du}\ e^{i u_2 L}
\ ,
&
\hat K(u)
=&\
\hat K^{(0)}(u)+\hat B_N(u)
\ ,
\label{eq:ON-icMERA}
\\
\cU(u_1,u_2)
=&\
e^{-iu_1 L} \ {\cal P} e^{-i \int_{u_2}^{u_1} \hat \cK(u) du}\ e^{i u_2 L}
\ ,
&
\hat \cK(u)
=&\
\hat \cK^{(0)}(u)+\hat \cB_N(u)
\ ,
\label{eq:ON-1-icMERA}
\end{align}
where $U$ preserves the $O(N)$ symmetry and $\cU$ spontaneously breaks it to $O(N-1)$.

For completess, a detailed description of the optimization procedure and equations for the icMERA tensor network for the $O(N)$ model can be found in Appendix \ref{app:appendix_b}.

\section{Entanglement Entropy and $1/N$ Expansion}
\label{sec:ON-ee}

In this section we show the results for the entanglement entropy of the half space in the scalar $O(N)$ model. The calculation is based on the same idea developed in Section \ref{sec:scalar}: when considering the ground state of a free theory, the entanglement entropy consists of the sum of the bond dimension for each field. 
Namely, the $O(N)$ icMERA circuit introduced above, generates non-Gaussian trial states which can be regarded as Gaussian wavefunctionals in a particular nonlinearly deformed field configuration basis.

As it will be shown, as in the case of the single scalar field, the landscape of non-Gaussian transformations that we may consider with a \emph{good} large $N$ behavior can be easily analyzed in the perturbative regime which is an useful guide in order to pick an specific non-Gaussian entangler \cite{Fernandez-Melgarejo:2020fzw,Ritschel:1992vr}.

\subsection{Entanglement entropy of the free theory}

Let us firstly consider the free $O(N)$ model. In this case, the ground state is Gaussian and the entanglement entropy is
\begin{align}
S_A
=
\sum_{i=1}^N S_i
=
\frac{|A_\perp|}{6}\sum_{i=1}^N\int_\vk \log\expval{\phi_i(\vk)\phi_i(-\vk)}_{\Psi_\Lambda}
\ ,
\label{eq:ON-ee}
\end{align}
where 
\begin{align}
\Psi_\Lambda[\bsPhi]
=
\braket{\bsPhi}{\Psi_\Lambda}
\ ,
\qquad
\ket{\Psi_\Lambda}
=
U^{(0)}(0,u_{\text{IR}})
\ket\Omega
\ ,
\end{align}
with $\ket\Omega$ satisfying $L\ket\Omega=0$ and
\begin{align}
\left(
	\sqrt M(\phi_i(\vk)-\chi_{0,i})+\frac{i}{\sqrt M}\pi_i(\vk)
	\right) \ket\Omega
	= 0 
\ ,
\qquad
\quad
\chi_{0,i}=\expval{\phi_i}_\Omega
\ ,
\qquad
i=1,\ldots,N
\ .
\label{eq:OmegaON}
\end{align}

In addition, when $U^{(0)}(0,u)$ acts on the vector field component $\phi_i(\vk)$, we obtain the transformed field
\begin{align}
\phi_i(\vk)
\quad
\to
\quad
\tilde \phi_i(\vk;u)
\equiv
[U^{(0)}(0,u)]^\dagger
\phi_i(\vk)
U^{(0)}(0,u)
\ .
\end{align}
Because of the commutation relations among the $N$ fields, this reduces to
\begin{align}
\tilde \phi_i(\vk;u)
=
[U_i^{(0)}(0,u)]^\dagger
\phi_i(\vk)
U_i^{(0)}(0,u)
\ ,
\quad
U_i^{(0)}(u_1,u_2)
\equiv
e^{-iu_1 L} \ {\cal P} e^{-i \int_{u_2}^{u_1} \hat K_i^{(0)}(u)du}\ e^{i u_2 L}
\ ,
\end{align}
where
\begin{align}
\hat K_i^{(0)}(u)
=
\frac12 \int_\vk \ g_i(k e^{-u};u)\left(
	\phi_i(\vk)\pi_i(-\vk)+\pi_i(\vk)\phi_i(-\vk)
	\right)
\ , 
\end{align}
with $i=1,\ldots,N$ (resp. $i=R,T$) for the case of the entangler \eqref{eq:entangler-ON-G} (resp. \eqref{eq:entangler-ON-G-RT}).

For later convenience, let us distinguish various types of wavefunctionals, depending on the basis of (un)transformed fields that we project on. When one single field is transformed $\phi_i\to\tilde \phi_i$, we denote the wavefunctional as
\begin{align}
\Psi[\tilde\phi_i;u]
&\equiv
\braket{\phi_1,\ldots,\phi_{i-1},\tilde\phi_i(u),\phi_{i+1},\ldots,\phi_N}{\Omega}\\ \nonumber
&=
\matrixel{\phi_1,\ldots,\phi_{i-1},\phi_i,\phi_{i+1},\ldots,\phi_N}{U_i^{(0)}(0,u)}{\Omega}
\ .
\end{align}
In contrast, if the state is projected over the $\tilde\phi_i$-basis, \emph{i.e.}, over the basis spanned by all the transformed fields, the wavefunctional is denoted by
\begin{align}
\Psi[\tilde\phi;u]
\equiv
\braket{\tilde \bsPhi(u)}{\Omega}
\equiv
\braket{\tilde \phi_1(u),\ldots,\tilde \phi_N(u)}{\Omega}
=
\matrixel{\phi_1,\ldots, \phi_N}{U^{(0)}(0,u)}{\Omega}
\ .
\end{align}
Then, the expression  \eqref{eq:ON-ee} for the entanglement entropy results \cite{Fernandez-Melgarejo:2021ymz}
\begin{align}
S_A
=&\
\cC_{d-1}\frac{|A_\perp|}{\epsilon^{d-1}}\sum_{i=1}^N\int_{u_{\text{IR}}}^0 du\ g_{i}(u) e^{u(d-1)}
+\text{const}'
\\
=&\
\cC_{d-1}\frac{|A_\perp|}{\epsilon^{d-1}}\sum_{i=1}^N\int_{u_{\text{IR}}}^0 du\ \sqrt{g^{i}_{uu}}\, e^{u(d-1)}
+\text{const}'
\ ,
\end{align}
where the Fisher information metric $g_{uu}^i$ is given by
\begin{align}
g^{i}_{uu}du^2
=
\cN^{-1}\left(
	1-\left|	\braket{\Psi_i(u)}{\Psi_i(u+du)}	\right|^2
	\right)
\ ,
\quad {\rm with} \quad
\ket{\Psi_i(u)}
=
U_i^{(0)}(0,u)\ket\Omega
\ .
\end{align}
In case the vacuum preserves $O(N)$, we have $g_i(u)\equiv g(u)$, $\forall i=1,\ldots, N$ and
\begin{align}
S_A
= N\ \cC_{d-1}\frac{|A_\perp|}{\epsilon^{d-1}}\int_{u_{\text{IR}}}^0 du\ g(u)\,  e^{u(d-1)}
+\text{const}'
\ .
\label{eq:ee-Gaussian-ON}
\end{align}
If we assume the aforementioned symmetry breaking  to $O(N-1)$, the entanglement entropy results
\begin{dmath}
S_A
= 
\cC_{d-1}\frac{|A_\perp|}{\epsilon^{d-1}}\int_{u_{\text{IR}}}^0 du\ \left[
	g_{R}(u)
	+(N-1) g_{T}(u) 
	\right]\, e^{u(d-1)}
+\text{const}'
= 
\cC_{d-1}\frac{|A_\perp|}{\epsilon^{d-1}}\int_{u_{\text{IR}}}^0 du\ \left[
	\sqrt{g^{R}_{uu}}
	+(N-1) \sqrt{g^{T}_{uu}}
	\right]\, e^{u(d-1)}
+\text{const}'
\ ,
\label{eq:ee-Gaussian-ONm1}
\end{dmath}
where the radial and transverse Fisher metrics are given by
\begin{align}
g^R_{uu}du^2
=&\
\cN^{-1}\left(
	1-\left|	\braket{\Psi_R(u)}{ \Psi_R(u+du)}	\right|^2
	\right)
\ ,
&
\ket{\Psi_R(u)}
=&\
U_R^{(0)}(0,u)\ket\Omega
\ ,
\\
g^T_{uu}du^2
=&\
\cN^{-1}\left(
	1-\left|	\braket{\Psi_T(u)}{ \Psi_T(u+du)}	\right|^2
	\right)
\ ,
&
\ket{ \Psi_T(u)}
=&\
U_T^{(0)}(0,u)\ket\Omega
\ ,
\end{align}
with the unitaries
\begin{align}
U_R^{(0)}(u_1,u_2)
\equiv &\
e^{-iu_1 L} \ {\cal P} e^{-i \int_{u_2}^{u_1} \hat K_R^{(0)}(u) du}\ e^{i u_2 L}
\ ,
&
U_T^{(0)}(u_1,u_2)
\equiv &\
e^{-iu_1 L} \ {\cal P} e^{-i \int_{u_2}^{u_1} \hat K_T^{(0)}(u) du}\ e^{i u_2 L}
\ .
\end{align}
As expected, in the limit $\chi_0\to0$, the $O(N)$ symmetry is restored and \eqref{eq:ee-Gaussian-ONm1} reduces to \eqref{eq:ee-Gaussian-ON}.

In both cases, the entanglement entropy consists of the sum of $N$ non-interacting scalar fields and is proportional to $N$.

\subsection{Entanglement Entropy of interacting $O(N)$ model}

In this subsection we are going to calculate the entanglement entropy of the interacting $O(N)$ model by summing the entanglement entropy of each of the $N$ fields at an arbitrary coupling $\lambda$. For this purpose, we are going to use the $O(N)$ icMERA circuit to generate non-Gaussian trial states. These states, which are genuinely non-Gaussian in the field basis $\{\phi_i\}$ in which the theory \eqref{eq:ON_theory} is expressed, turn out to be Gaussian in a basis of fields $\{\tilde\phi_i\}$. The transformation that relates both bases is precisely the $O(N)$ icMERA circuit that we have introduced above.

We will consider both the unitaries containing the $O(N-1)$ entangler \eqref{eq:ON-icMERA} and the $O(N)$ entangler \eqref{eq:ON-icMERA}.

\subsubsection{$O(N-1)$ icMERA}
\label{sec:ON-1-ee-exact}

When applying the $O(N-1)$ icMERA circuit $\cU(0,u)$, the vector field operator $\bsPhi$ gets transformed as
\begin{align}
\tilde\bsPhi(\vk;u)=
\cU^\dagger(0,u)\, 
\bsPhi(\vk)\, 
\cU(0,u)
\ . 
\end{align}
This implies that 
\begin{align}
\tilde\Psi[\bsPhi;u]
\equiv
\Psi[\tilde\bsPhi;u]
=
\Psi[\bsPhi-s\ \bsPhi \cdot\bsPhi;\ u]
\ ,
\end{align}
where
$\Psi[\tilde\bsPhi;u]$ is a Gaussian wavefunctional whose argument has been nonlinearly displaced $\bsPhi\to\bsPhi-s\ \bsPhi \cdot\bsPhi$ \cite{IbanezMeier:1991hm}.

As a consequence, the entanglement entropy for the non-Gaussian wavefunctional $\tilde\Psi_\Lambda[\bsPhi]=\Psi_\Lambda[\tilde\bsPhi]=\braket{\tilde\bsPhi}{\Psi_\Lambda}$ can be obtained using the same expression as for a Gaussian state (\emph{cf.} \eqref{eq:ee-Gaussian-ON}--\eqref{eq:ee-Gaussian-ONm1}), but just replacing the fields $\phi(\vp)$ by the transformed fields $\tilde\phi(\vp)$,
\begin{align}
S_A^{\Psi[\tilde\bsPhi]}
=
\frac{|A_\perp|}{6}
\sum_{i=1}^N
\int_\vp\log\expval{\tilde\phi_i(\vp)\tilde\phi_i(-\vp)}_{\Psi_\Lambda[\tilde\bsPhi]}
\ .
\label{eq:ON-ee-exact}
\end{align}
Precisely, and reasoning along the lines of Section \ref{sec:scalar} (see also \cite{Fernandez-Melgarejo:2020utg}), because $\Psi_\Lambda[\tilde\bsPhi;u]$ is a genuine Gaussian state in the $\tilde\bsPhi$-basis, there exists an $O(N)$ Gaussian cMERA circuit defined by an entangler
\begin{align}
\hat{\tilde \cK}_N^{(0)}(u)
=
\frac12 \int_\vk \tilde g(k e^{-u};u)\left[ \tilde\bsPi(\vk)\cdot\tilde\bsPhi(-\vk)+\tilde\bsPhi(\vk)\cdot\tilde\bsPi(-\vk)\right]
\ ,
\qquad
\tilde g(k;u)
\equiv
\tilde g(u)\cdot\Gamma(k/\Lambda)
\ ,
\end{align}
which generates such state (for a particular variational parameter) when acting on $\ket\Omega$.
Then, according to the above result in Eq. \eqref{eq:ee-Gaussian-ONm1}, which entails a generalization of the expression for a single field \cite{Fernandez-Melgarejo:2021ymz}, the half space entanglement entropy associated to this state is
\begin{dmath}
S_A
= \sum_{i=1}^N\ \cC_{d-1}\frac{|A_\perp|}{\epsilon^{d-1}}\int_{u_{\text{IR}}}^0 du\ \tilde g_{i}(u)\, e^{u(d-1)}
+\text{const}'
= \sum_{i=1}^N\ \cC_{d-1}\frac{|A_\perp|}{\epsilon^{d-1}}\int_{u_{\text{IR}}}^0 du\  \sqrt{\tilde g^i_{uu}}\, e^{u(d-1)}
+\text{const}'
\ .
\label{eq:ee-Gaussian-ON-tilde}
\end{dmath}
Upon the identification $\tilde\bsPhi=(\tilde \phi_R,\vec{\tilde\phi}_T)$, this expression results
\begin{align}
S_A
= 
\cC_{d-1}\frac{|A_\perp|}{\epsilon^{d-1}}\int_{u_{\text{IR}}}^0 du\ \left(
	\sqrt{\tilde g^{R}_{uu}}
	+(N-1) \sqrt{\tilde g^{T}_{uu}}
	\right)e^{u(d-1)}
+\text{const}'
\ .
\end{align} 
Let us note that each of the Fisher metrics in these expressions are calculated from the fidelity of the different states, which are specified here:
\begin{align}
\tilde g^R_{uu}du^2
=&\
\cN^{-1}\left(
	1-\left|	\braket{\tilde \Psi_R(u)}{\tilde \Psi_R(u+du)}	\right|^2
	\right)
\ ,
&
\ket{\tilde \Psi_R(u)}
=&\
U_R(0,u)\ket\Omega
\ ,
\\
\tilde g^T_{uu}du^2
=&\
\cN^{-1}\left(
	1-\left|	\braket{\tilde \Psi_T(u)}{\tilde \Psi_T(u+du)}	\right|^2
	\right)
\ ,
&
\ket{\tilde \Psi_T(u)}
=&\
U_T(0,u)\ket\Omega
\ ,
\label{eq:fidelities_rt}
\end{align}
where
\begin{align}
U_R(u_1,u_2)
\equiv &\
e^{-iu_1 L} \ {\cal P} e^{-i \int_{u_2}^{u_1} \left( \hat K_R^{(0)}(u)+\hat \cB_N(u)\right) du}\ e^{i u_2 L}
\ ,
\\
U_T(u_1,u_2)
\equiv &\
e^{-iu_1 L} \ {\cal P} e^{-i \int_{u_2}^{u_1} \hat K_T^{(0)}(u) du}\ e^{i u_2 L}
\ .
\end{align}
In analogy with the results for the single scalar in Sec. \ref{sec:scalar-ee}, the entangler $\hat{\tilde{\cK}}_N^{(0)}$ is solely used to express the entanglement entropy in terms of Fisher metrics associated to Gaussian states in the $\tilde\phi_i$-basis. However, to evaluate these quantities we do not need to obtain $\tilde g(u)$, the variational parameter in $\hat{\tilde{\cK}}_N^{(0)}$, from the minimization of a parent Hamiltonian $\cH[\tilde\bsPhi]$. Instead, the Fisher metric enables us to associate $\tilde g(u)$ with the icMERA variational parameters in $\hat \cK(u)$, Eq. \eqref{eq:ON-icMERA}, which are obtained from the minimization of the interacting energy functional.

The Fisher metrics $\tilde g_{uu}^R$ and $\tilde g_{uu}^T$, which are essential to obtain explicit expressions of the entanglement entropy, are given by
\begin{align}
\tilde g_{uu}^R
=&\
\cN^{-1}\left(\expval{\left( \hat K_R^{(0)}(u)+\hat{\cB}_N(u)\right)^2 }_\Omega
-\expval{\hat K_R^{(0)}(u)+\hat \cB_N(u)}^2_\Omega\right)
\ ,
\\
\tilde g_{uu}^T
=&\
\cN^{-1}\left(\expval{\hat K_T^{(0)}(u)^2 }_\Omega
-\expval{\hat K_T^{(0)}(u)}^2_\Omega\right)
\ .
\end{align}
When evaluated, they result
\begin{align}
\begin{split}
\tilde g_{uu}^R
=&\
g_R(u)^2 + \gamma^2\, e^{-u(d-1)} \left[\phi_c^2
+2\, (N-1) \bar \chi_2(u)\right]
\ ,
\\
\tilde g_{uu}^T
=&\
g_T(u)^2
\ ,
\end{split}
\end{align}
where $\phi_c$, which accounts for the non-Gaussian disconnected 2-point correlators, is given by
\begin{align*}
\phi_c =  (N-1)\, \bar \chi_1(u)
\, .
\end{align*}  and $\bar \chi_2(u)$  is given in \eqref{eq:chi1chi2}.\footnote{The explicit evaluation of these integrals is provided in Appendix \ref{app:appendix}.}
Masses, cutoffs and the coupling are contained in the following parameters:
\begin{equation}
M_R = \sqrt{\Lambda^2 + \mu_R^2}\, , \quad M_T = \sqrt{\Lambda^2 + \mu_T^2}\, , \quad \gamma^2 = \frac{\beta^2}{2\, \Lambda^{(d-1)}}\frac{M_R}{M_T^2}\, .
\end{equation}

\begin{figure}[!t]
\centering
\includegraphics[width=\textwidth]{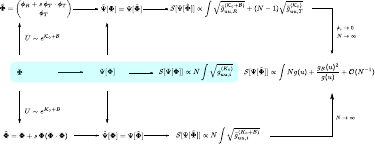}
\caption{
\textit{
Schematic relation among Gaussian and non-Gaussian objects. \emph{Central row:} for a field $\bsPhi$ with a Gaussian wavefunctional $\Psi[\bsPhi]$, the entanglement entropy is given by $S[\Psi[\bsPhi]]$. \emph{Top row:} for the transformed field $\tilde\bsPhi$ (radial and transverse fields break $O(N)$ to $O(N-1)$), the non-Gaussian wavefunctional $\tilde\Psi[\bsPhi]$ can be written as a Gaussian state in the $\tilde\bsPhi$-basis. Consequently, the entanglement entropy is given by $S[\Psi[\tilde\bsPhi]]$. \emph{Bottom row:} assuming the transformed field $\tilde\bsPhi$ ($O(N)$ is preserved), the non-Gaussian wavefunctional $\tilde\Psi[\bsPhi]$ can be written as a Gaussian state in the $\tilde\bsPhi$-basis. Consequently, the entanglement entropy is given by $S[\Psi[\tilde\bsPhi]]$. Up to $\cO(N^{-1})$, both entanglement entropies coincide for $N\to\infty$.
}}
\label{fig:scheme2 }
\end{figure}

With this, then the entanglement entropy can be written as 
\begin{align}
S_A
= 
\cC_{d-1}\frac{|A_\perp|}{\epsilon^{d-1}}\int_{u_{\text{IR}}}^0 du\, e^{u(d-1)}\, s(u)
+\text{const}'
\ ,
\label{eq:ee-ON-final}
\end{align}
where $s(u)=(N-1)s_T(u)+s_R(u)$ is given by
\begin{align}
s_T(u)
= g_{T}(u)  
\ ,
\qquad
s_R(u)
= \sqrt{g_R(u)^2
+\gamma^2\, e^{-u(d-1)} \left[
	\phi_c^2
	+ 2(N-1)\,  \bar \chi_2(u)
	\right]}
\ .
\label{eq:su_ee-ON-final}
\end{align}
We remark again that this result is obtained solely by considering the Gaussian nature of the icMERA wavefunctionals in the $\tilde \phi$-basis and the results for the half space entropy in a Gaussian cMERA \cite{Fernandez-Melgarejo:2021ymz}. Similar to the single self interacting scalar field, Eq. \eqref{eq:ee-ON-final} suggests that $s(u)$ is a local density of disentanglers, \emph{a.k.a.} bond dimension, defined in terms of a tensor network-like Fisher metric of the $O(N)$ icMERA circuit.
In the next subsection we will compare to what extent this bond dimension correctly captures the entanglement entropy given by \eqref{eq:su_ee-ON-final}. 

It is also worth to mention that  \eqref{eq:su_ee-ON-final} is valid for any value of the interaction strength and for any finite value $N$. The limit $\beta\to 0$, for which the non-Gaussian effects vanish, is a necessary condition to restore the $O(N)$ symmetry, though not sufficient. For this to occur, we also have to impose $g_R(u)=g_T(u)$.

Our interest now is to obtain the large $N$ limit of the expressions for the entanglement entropy. Following \cite{Stevenson:1986nb}, we take the limit $N \to \infty$ with
\begin{equation}
\bar \lambda =  N\, \lambda\, , \quad {\bar \phi_c}^2=\frac{\phi_c^2}{N}\, ,
\label{eq:largeN}
\end{equation}
held constant. This is the scaling that allows the Gaussian effective potential to be of order $N$. This is relevant as far as the optimization equations for the icMERA circuits under consideration yield \cite{IbanezMeier:1991hm}
\begin{equation}
\beta = -4\lambda\, \phi_c = 
-4\, \bar \lambda\,  {\bar \phi_c}\, N^{-1/2} = \bar \beta \, N^{-1/2}\, ,
\end{equation}
for the icMERA circuit with a non-Gaussian entangler \eqref{eq:entangler-ON-NG}. This implies that $\beta^2 = \bar \beta^2/N$, with $\bar \beta$ held fixed.


%
Then, when taking $N\to\infty$ for $s(u)$, we have
\begin{align}\label{eq:su_largeN}
s(u)
=
N\, g_T(u)
+\sqrt{g_R(u)^2+\bar\gamma^2\, e^{-u(d-1)} \left(2\, \bar\chi_2(u)+ \bar \phi_c^2\right)}
+\cO(1/N)
\ .
\end{align}
and consequently, the entropy reads
\begin{align}
S_A 
=&\ N\, \cC_{d-1}\frac{|A_\perp|}{\epsilon^{d-1}}\int_{u_{\text{IR}}}^0 du\ g_T(u) e^{u(d-1)}
+\text{const}'\\ \nonumber
&\ + \cC_{d-1}\frac{|A_\perp|}{\epsilon^{d-1}}\int_{u_{\text{IR}}}^0 du\, \sqrt{g_R(u)^2+\bar\gamma^2\, e^{-u(d-1)} \left(2\,\bar \chi_2(u)+ \bar \phi_c^2\right)}\, e^{u(d-1)}
+\cO(N^{-1})
\ .
\end{align}
Let us analyze this result in detail. We note that the structure of the entanglement entropy in the large $N$ limit is such that, first, there is a leading $\cO(N)$ term $\propto g_{T}(u)$ arising from the $N-1$ transverse fields. This is the leading term in the entanglement entropy that can be related with the RT minimal surface in cMERA \cite{Fernandez-Melgarejo:2021ymz}. On the other hand, there is a subleading term $\cO(N^0)$, depending on $g_R(u)$ and $\bar\chi_2(u)$ which, as in the case of the single scalar field, can be interpreted as the ``connected'' quantum correction to the entanglement entropy. Let us also mention that, as $\bar\chi_2(u)\propto g_B(u)^2$ the density $s(u)$ is a homogeneous polynomial of degree 1 with respect to variational parameters.

\subsubsection{$O(N)$ icMERA}
\label{sec:ON-ee-exact}

In this section we will consider the unitary $U(u_1,u_2)$ given in Eq. \eqref{eq:ON-icMERA}, which contains the $O(N)$ symmetric entangler $B_N(u)$ of Eq. \eqref{eq:entangler-ON-G-sym}.

Firstly,  we are going to calculate the entanglement entropy by considering the sum of the entanglement entropy of each of the $N$ fields $\tilde\phi_i$ by using the expression \eqref{eq:ON-ee-exact}. 
In this case, each field $\tilde\phi_i$ is the byproduct of the icMERA circuit through the factorization
\begin{align}
U_i(u_1,u_2)
\equiv
e^{-iu_1L}
\cP
e^{-i\int_{u_1}^{u_2}\left(
	\hat K_i^{(0)}(u)
	+\beta_N \hat B_N(u)
	\right)du}
e^{iu_2L}
\ ,
\label{eq:ON-unitary-exact}
\end{align}
where $\beta_N = \beta\, N^{-\alpha}$ and $\alpha$ is a free parameter that will be justified below. 

Because the $N$ fields $\phi_i$ are equally transformed, the entanglement entropy in \eqref{eq:ee-Gaussian-ON-tilde} results
\begin{align}
S_A
=
N \cC_{d-1}\frac{|A_\perp|}{\epsilon^{d-1}}
\int_{u_{\text{IR}}}^0 du \sqrt{\tilde g_{uu}^i(u)}e^{u(d-1)} + {\rm const}'
 \ ,
\label{eq:ON-ee-symmetric}
\end{align}
where the Fisher metric $\tilde g_{uu}^i$ is, $\forall\, i=1 \cdots N$,
\begin{align}
\tilde g^i_{uu}du^2
=&\
\cN^{-1}\left(
	1-\left|	\braket{\tilde \Psi_i(u)}{\tilde \Psi_i(u+du)}	\right|^2
	\right)	
\ ,
&
\ket{\tilde \Psi_i(u)}
=&\
U(0,u)\ket\Omega
\ .
\label{eq:ON-fidelity-sym-2}
\end{align}
Let us explain the parameter $\beta_N$. The Fisher metric \eqref{eq:ON-fidelity-sym-2} has been calculated for the state $\ket{\tilde \Psi_i(u)}$, which is the one obtained upon the transformation of just one single field $\phi_i\to\tilde\phi_i$. Namely, the entanglement entropy \eqref{eq:ON-ee-symmetric} consists of the summation of $N$ bond dimensions, each of them corresponding to the non-linear transformation of a single field $\phi_i$, while the rest remain unchanged.
Consequently, in order for the Gaussian contribution to be the leading $\cO(N^0)$ term, we take $\alpha=1/2$. Equivalently, this can be understood as the relativative normalization factor between $\ket{\tilde\Psi_i(u)}$ in \eqref{eq:ON-fidelity-sym-2} and $\ket{\tilde\Psi(u)}=U(0,u)\ket\Omega$.

When explicitly calculated, the Fisher information metric results
\begin{align}
\tilde g_{uu}^i(u)
=
g(u)^2
+\frac{\beta^2}{4 M^2\Lambda^{(d-1)}}\, e^{-u(d-1)}\frac{(N^2+2N)}{N}\, \bar\zeta(u)
\ ,
\end{align}
where $\bar\zeta(u)$ is the loop integral which is related to the connected part of the 2-point correlator,
\begin{align}
\bar\zeta(u)=&\
g_B(u)^2\cdot I_3(\Lambda e^u)
\ ,
\\
I_3(\Lambda e^u)
\equiv
&\
\f14 \int_{\vp \vq \vr}\, \Gamma(|\vp+\vq+\vr|e^{-u}/\Lambda)\Gamma(p e^{-u}/\Lambda)\, \Gamma(q e^{-u}/\Lambda)\, \Gamma(re^{-u}/\Lambda)
\ .
\label{eq:def_zeta_u}
\end{align}
Then, when plugged in \eqref{eq:ON-ee-symmetric}, we have
\begin{align}
S_A
=
N\, \cC_{d-1}\frac{|A_\perp|}{\epsilon^{d-1}}
\int_{u_{\text{IR}}}^0 du \ e^{u(d-1)} \sqrt{
g(u)^2
+\frac{\beta^2}{ 4 M^2\Lambda^{(d-1)}}\, e^{-u(d-1)}(N+2)\bar\zeta(u)}
 \ ,
\label{eq:ON-ee-symmetric-final}
\end{align}

Let us remark that the $N$ dependence of the Fisher metric-like-corrections to the entanglement entropy is similar to the perturbative two-loop corrections to entanglement entropy in the $O(N)$ model obtained in \cite{Chen:2014wka}. Namely, the Feynman diagram involved in this two-loop calculation, is the first one in the series of diagrams generated by the Schwinger-Dyson equation appearing in the optimization of the icMERA circuit (see Appendix \ref{app:appendix_b}, \cite{Ritschel:1992vr,Fernandez-Melgarejo:2019sjo})

To analyze the $1/N$ expansion for this transformation, let us note the scaling of the variational parameter $s$ is given by (\emph{cf.} \eqref{eq:largeN})
\begin{equation}
\beta = -4\lambda = -4\, \bar \lambda\, N^{-1} = \bar \beta \, N^{-1}\, ,
\end{equation}
such that $\beta^2 = \bar \beta^2/N^2$. As it is an $O(N)$ symmetry preserving transformation, there is no condensate $\phi_c$. Then, taking the $N \to \infty$ limit in \eqref{eq:ON-ee-symmetric-final}, and defining $\gamma^2 = \beta^2/8 M^2 \Lambda^{(d-1)}$, we have
\begin{align}
\begin{split}
S_A
=&\ 
N \cC_{d-1}\frac{|A_\perp|}{\epsilon^{d-1}}\int_{u_{\text{IR}}}^0 du \ e^{u(d-1)} \sqrt{g(u)^2+ \frac{2\, \gamma^2}{N^2}\, e^{-u(d-1)}\, (N+2)\bar\zeta(u)}
\\
=&\
\cC_{d-1}\frac{|A_\perp|}{\epsilon^{d-1}}\int_{u_{\text{IR}}}^0 du \ e^{u(d-1)} \left(
	N g(u) 
	+\bar\gamma^2\, e^{-u(d-1)} \left(\frac{\bar \zeta(u)}{g(u)} \right) \right)+\cO(N^{-1})
\\
=&
N\, \cC_{d-1}\frac{|A_\perp|}{\epsilon^{d-1}}\int_{u_{\text{IR}}}^0 du\, g(u)\,  e^{u(d-1)}  
	+\cC_{d-1} |A_\perp|\, \frac{\bar \beta^2}{8 M^2} \int_{u_{\text{IR}}}^0 du \left(\frac{\bar \zeta(u)}{g(u)} \right) +\cO(N^{-1})	
\ .
\end{split}
\label{eq:ee-ON-largeN}
\end{align}

To check the consistency of these results, in the next subsection we are going to calculate the entanglement entropy by exclusively using the effective bond dimension of the $O(N)$ icMERA circuits.

\subsection{Effective bond dimension in icMERA}

Having obtained the entanglement entropy of the $O(N)$ model from the 2-point correlators evaluated with the icMERA Gaussian wavefunctionals (in the deformed basis of the transformed fields $\tilde \phi$), our aim here is 
to calculate the entanglement entropy of the $O(N)$ model by exclusively integrating the bond dimension, \emph{i.e.}, the Fisher information metric of the icMERA circuit that is defined through the unitaries $U$ and $\cU$ given in \eqref{eq:ON-icMERA} and \eqref{eq:ON-1-icMERA}. This derivation entails a generalization of the entanglement entropy for the half space in the free theory given in \cite{Fernandez-Melgarejo:2021ymz} for one single scalar and \eqref{eq:ON-ee} for the $O(N)$ model:
\begin{align}
S_A
= 
N\, \cC_{d-1}\frac{|A_\perp|}{\epsilon^{d-1}}\int_{u_{\text{IR}}}^0 du\, g_{\rm eff}(u)\, e^{u(d-1)}
+\text{const}'
\ ,
\label{eq:S-eff}
\end{align}

In this sense, despite the QFT origin of the ``effective variational parameter'' $g_{\rm eff}(u)$ is not clarified yet, we will show that, in the large $N$ limit, this formula provides a correct prediction for the entanglement entropy when substituting $g_{\text{eff}}(u)$ by the Fisher information metric $g^{\rm eff}_{uu} $ of the icMERA circuit generated by $\hat K(u)$,
\begin{align}
 g^{\text{eff}}_{uu}\, du^2=  g_{\text{eff}}(u)^2du^2
=&\
\cN_N^{-1}\left(
	1-\left|	\braket{\tilde \Psi(u)}{\tilde \Psi(u+du)}	\right|^2
	\right)
\ ,
\label{eq:1-fidelity-ON}
\end{align}
with $|\tilde \Psi(u)\rangle=U(0,u)\ket\Omega$ and
\begin{align}
U(u_1,u_2)
& \equiv 
e^{-iu_1 L} \ {\cal P} e^{-i \int_{u_2}^{u_1} \left( \hat K^{(0)}(u)+\hat B_N(u)\right) du}\ e^{i u_2 L}\, .
\end{align}

\subsubsection*{$O(N-1)$ Bond Dimension}

In order to compare these results with those obtained in Section \ref{sec:ON-1-ee-exact}, we note that $\ket{\tilde\Psi(u)}\propto\sqrt N\ket{\Psi_{R,T}(u)}$. This results in a relation between the normalization constants: $\cN_N = N \cdot\cN $.\footnote{This mismatch between the Fisher information metric and the entanglement entropy reduces to an overall coefficients, which was suggested in \cite{Swingle:2012wq}, is unambiguously fixed by the Gaussian term, which has to be $\cO(N)$.}
The result is
\begin{dmath}
g_{uu}^{\rm eff}(u)
=\frac{(N-1)}{N}\,  g_T(u)^2
+ \frac
{g_R(u)^2}{N}
+\frac{\gamma^2}{N} e^{-u(d-1)}\left[
	\phi_c^2\,
	+ 2(N-1)\,  \bar \chi_2(u)
	\right]
\ .
\label{eq:icMERA_metric_ON_a}
\end{dmath}
Plugging this expression into the entanglement entropy formula \eqref{eq:S-eff}, we have
\begin{multline}
S_A
= 
N\, \cC_{d-1}\frac{|A_\perp|}{\epsilon^{d-1}}\int_{u_{\text{IR}}}^0 du\, e^{u(d-1)} 
	\Bigg(\frac{(N-1)}{N}\,  g_T(u)^2
+ \frac
{g_R(u)^2}{N}
+\frac{\gamma^2}{N}\, e^{-u(d-1)}\left[
	\phi_c^2\,
	+ 2(N-1)\,  \bar \chi_2(u)
	\right]\Bigg)^{1/2}
\ .
\end{multline}
This result for finite $N$ is qualitatively similar to \eqref{eq:ee-ON-final} but not exactly the same. Taking $N \to \infty$ in the effective bond dimension \eqref{eq:icMERA_metric_ON_a},   we obtain
\begin{align}
 g_{uu}^{\rm eff}(u)
 &=
g_T(u)^2 
+\frac{1}{N}\left[g_R(u)^2 + \bar \gamma^2 e^{-u(d-1)} \left(2\, \bar \chi_2(u) + \bar \phi_c^2\right)\right]
\ .
\label{eq:icMERA_metric_ON_a2}
\end{align}
From this, we read off the \emph{effective} disentangler strength as
\begin{align}
g_{\rm eff}(u)
=&\ g_{T}(u)\, 
+\frac{1}{N}\, g_{R,\chi}(u)\, 
\ ,
\qquad
g_{R,\chi}(u) =
\left(\frac{g_R(u)^2 + \bar \gamma^2 e^{-u(d-1)} \left(2\, \bar \chi_2(u) + \bar \phi_c^2\right)}{2\, g_{T}(u)}\right)\, 
\, .
\label{eq:icMERA_eff_bd}
\end{align}
Again, we observe that the structure of the effective bond dimension in the large $N$ limit is such that there is a leading $\cO(1)$ term $g_{T}(u)$ arising from the $N-1$ transverse fields in the Gaussian part of the entangler. This contribution, as it will be shown in Section \ref{sec:holo_icmera}, is the leading term in the entanglement entropy that can be related with the RT minimal surface in cMERA \cite{Fernandez-Melgarejo:2021ymz}. On the other hand, there is a subleading term $\cO(1/N)$, depending on $\bar\chi_2(u)$, which, as in the case of the single scalar field, can be interpreted as the ``connected'' contribution to the \emph{effective} bond dimension  \cite{Fernandez-Melgarejo:2020fzw}.  

Thus, according to \cite{Fernandez-Melgarejo:2021ymz} and Section \ref{sec:scalar}, the entanglement entropy calculated through this local bond dimension in cMERA will inherit the property that, at large $N$, the ``connected'' quantum fluctuations are strongly suppressed. To see this, we recall that our prescription to compute the entanglement entropy for the half space is given in \eqref{eq:S-eff} \cite{Fernandez-Melgarejo:2021ymz}.

With this we may use \eqref{eq:icMERA_eff_bd}, to obtain
\begin{align}\label{eq:largeN_ee_chi}
S_A 
=&\ 
S_{T}^{(N)} + \tilde S_{R,\chi}^{(1)}
\, ,
\end{align}
where
\begin{align}\label{eq:largeN_ee_leading}
S_T^{(N)} \equiv  N\, \cC_{d-1}\frac{|A_\perp|}{\epsilon^{d-1}}\int_{u_{\text{IR}}}^0 du\, g_{T}(u)\, e^{u(d-1)}+\text{const}' 
\, ,
\end{align}
is the entropy of $N$ free fields with a bond dimension $g_{T}(u)$ and\footnote{See Appendix \ref{app:appendix}.}
\begin{align*}
\tilde S_{R,\chi}^{(1)}\equiv \cC_{d-1}\frac{|A_\perp|}{\epsilon^{d-1}}\int_{u_{\text{IR}}}^0 du\, g_{R,\chi}(u)\, e^{u(d-1)}
\, ,
\end{align*}
amounts to the $\cO(N^0)$ quantum correction induced by the non-Gaussian term in the icMERA entangler. 

Here we note the following. As it is shown in Appendix \ref{app:appendix_b}, the icMERA optimization is usually carried out for $\bar \phi_c \sim 0$. When this occurs, the $O(N)$ symmetry is restored and $g_R(u) = g_T(u)\equiv g(u)$ as commented above. Interestingly, in this regime, $g_{\rm eff}(u)$ in \eqref{eq:icMERA_eff_bd} and the entanglement density \emph{per scale} $s(u)$ in \eqref{eq:su_largeN} are related by
\begin{align}
g_{\rm eff}(u) \sim \frac{s(u)}{N} \approx g(u) + \frac{\bar\gamma^2}{N} \, e^{-u(d-1)}\, \left(\frac{\bar \chi_2(u)}{g(u)}\right) 
\, ,
\end{align}
which yields
\begin{align}
S^{(N)} &=  N\, \cC_{d-1}\frac{|A_\perp|}{\epsilon^{d-1}}\int_{u_{\text{IR}}}^0 du\, g(u)\, e^{u(d-1)}+\text{const}' 
\\
\tilde S_{\chi}^{(1)}&=\cC_{d-1}\frac{\bar\beta^2}{2 M}|A_\perp|\int_{u_{\text{IR}}}^0 du\, \left(\frac{\bar \chi_2(u)}{g(u)} \right)
\, .
\end{align}

%
%

\subsubsection*{$O(N)$ Bond Dimension}

Let us obtain now the bond dimension associated to the $O(N)$ unitary $U(u_1,u_2)$, \eqref{eq:ON-icMERA}.

Firstly, the effective Fisher metric of the icMERA circuit \eqref{eq:1-fidelity-ON} where $B_N$ is the entangler in \eqref{eq:entangler-ON-G-sym} results as
\begin{align}\label{eq:ON-fidelity-sym-1}
g_{uu}^{\rm eff}(u)
&=
 g(u)^2
+\frac{\beta ^2 }{16 M^2\Lambda^{(d-1)}}\, e^{-u(d-1)} \frac{(N^2+2N)}{N}\int_{\vp \vq \vr} c\left(\vp e^{-u},\vq e^{-u},\vr e^{-u};u\right)^2 \\ \nonumber
&=
g(u)^2
+\frac{\beta ^2 }{4 M^2\Lambda^{(d-1)}}\, e^{-u(d-1)}  (N+2)\, \bar \zeta(u)
\, ,
\qquad
M
=\sqrt{\Lambda^2+\mu^2}
\end{align}
where $\bar\zeta(u)$, which is related to the connected part of the 2-point correlator \cite{Fernandez-Melgarejo:2020fzw}, is given in \eqref{eq:def_zeta_u}.
We emphasize that this metric accounts for the distance between two states when the icMERA circuit \eqref{eq:ON-icMERA} is applied. In particular, this unitary operator transforms the $N$ fields $\bsPhi\to\tilde\bsPhi$. 

Let us remember that $N$-renormalized coupling is $\bar\beta=\beta N$. Consequently, the effective bond dimension $g_{\text{eff}}(u)=\sqrt{\tilde g_{uu}^{\text{eff}}(u)}$ from  \eqref{eq:ON-fidelity-sym-1} can be written as
\begin{align}
g_{\text{eff}}(u)^2
= &\
g(u)^2
+ 2\, \gamma^2 \, e^{-u(d-1)} \, \left(N+2\right) \, \bar \zeta(u)\, ,\quad \gamma^2 = \frac{\beta^2}{8 M^2\, \Lambda^{(d-1)}}
\ .
\end{align}
When taking the limit $N\to\infty$ we have
\begin{align}
g_{\text{eff}}(u) = g(u)+ \frac{\bar \gamma^2}{N}\, e^{-u(d-1)}\, \left(\frac{\bar \zeta(u)}{g(u)} \right) +\cO(N^{-2})
\, .
\end{align}
When plugging this expression in the entanglement entropy formula \eqref{eq:S-eff}, we have
\begin{align}\label{eq:ON-ee-largeN1}
S_A
=&\
N\, \cC_{d-1}\frac{|A_\perp|}{\epsilon^{d-1}}\int_{u_{\text{IR}}}^0 du\, g(u)\, e^{u(d-1)}+\text{const}' \\ \nonumber
&+ \cC_{d-1}|A_\perp|\, \frac{\bar\beta^2}{8 M^2}\, \int_{u_{\text{IR}}}^0 du\, \left(\frac{\bar \zeta(u)}{g(u)} \right) +\cO(1/N)\\ \nonumber
\, .
\end{align}
In this case, we find a perfect agreement with the result obtained from the sum of the entropies of each of the $N$ fields, Eq. \eqref{eq:ee-ON-largeN}.

\section{Holographic features of icMERA tensor networks}
\label{sec:holo_icmera}

In this section we will discuss the entanglement entropy in the large $N$ limit of  icMERA in the light of holography. 

The $1/N$ corrections at strong coupling of the Ryu-Takayanagi formula may modify the leading area term. However, these have not been generally understood and there are reasons to think that they are not connected to the area term. According to the results of the entanglement entropy in the $1/N$ expansion, the Gaussian term of the disentangler give an account (leading term, $\cO(N)$) of the RT prescription in holography \cite{Fernandez-Melgarejo:2021ymz}. In addition, the quantum corrections to the entanglement entropy due to interactions, are due to the  non-Gaussian part of the icMERA entangler. Those corrections, which are $\cO(1)$, are suppressed in the large $N$ limit, thus making the entropy to be always related to the area term, regardless of the coupling strength.

Let us firstly recapitulate our results by looking at the $1/N$ expansions of the $O(N)$ model entanglement entropy, \eqref{eq:largeN_ee_chi} and \eqref{eq:ON-ee-largeN1}. These expressions clearly show that the entanglement entropy computed through the icMERA bond dimension (quantum Fisher information metric), is given by the sum of two terms: a leading semiclassical contribution and a subleading quantum term. This interpretation comes by  simply keeping track of the orders $G_N^{-1}\, (c_{\rm eff}\sim N)$ and $G_N^0\, (c_{\rm eff}^0\sim N^0)$, respectively. Namely, our analysis shows that different choices of icMERA tensor networks that differ only at subleading orders in $1/N$ at the level of the effective bond dimension, converge up to subleading corrections. Here, we argue that these subleading quantum corrections to entanglement entropy in  icMERA, can be interpreted  as  quantum corrections to the RT semiclassical leading term in the holographic entanglement entropy.

To justify this, let us just rewrite \eqref{eq:flm} as
\begin{align}
\begin{split}
S_A = &\ \frac{\text{Area}(\gamma_A)}{4 G_N} + S_{\textrm q} + \cO(G_N) 
\\
\equiv&\ c_{{\text{eff}}}\, \expval{\widehat{\text{Area}}(\gamma_A)} + S_{\textrm q}  + \cO(c_{\text{eff}}^{-1})
\, .
\end{split}
\label{eq:flm_operator}
\end{align}
The first line in \eqref{eq:flm_operator} is stated in a semiclassical gravity approximation given in terms of geometric quantities and  quantum corrections computed at a fixed background. The second line in \eqref{eq:flm_operator} intends to formalize this statement in a quantum gravitational theory in the bulk. Indeed, as holography and particularly the AdS/CFT correspondence, establishes a duality between a boundary quantum field theory and a gravitational theory in the bulk, it is reasonable to relate operators pertaining to these two theories. In this sense, at tree level in the large $N$ limit, the quantum gravitational theory admits an $\widehat{\text{Area}}$  operator whose expectation value gives the classical area of the  RT minimal surface. Consequently, the leading terms in \eqref{eq:largeN_ee_chi} and \eqref{eq:ON-ee-largeN1}  correspond to the expectation value of the $\widehat{\text{Area}}$  operator on the RT surface $\gamma_A$.

To see this more explicitly, let us first consider \eqref{eq:largeN_ee_leading}. As commented above,
\footnote{We will generalize for $N$ fields the analysis done in \cite{Fernandez-Melgarejo:2021ymz}.}
\begin{equation}
S_T^{(N)} \equiv  N\, \cC_{d-1}\frac{|A_\perp|}{\epsilon^{d-1}}\int_{u_{\text{IR}}}^0 du\, g_{T}(u)\, e^{u(d-1)}+\text{const}' 
\, .
\end{equation}
We are going to define local areas in the bulk in terms of the differential entanglement generated along the tensor network, as it was suggested in \cite{Swingle:2012wq}. That is, we have
\begin{align}\label{eq:diff_ee_icmera}
\frac{d S_T^{(N)}}{du}
= 
N\, \cC_{d-1} \cdot d\mathbb{A}_{\text{TN}}
\, ,
\end{align} 
where
\begin{align}
d\mathbb{A}_{\text{TN}}\equiv \frac{|A_\perp|}{\epsilon^{d-1}}\, g_T(u)\,  e^{u(d-1)}\, du
\, ,
\label{eq:darea_cmera}
\end{align}
is the infinitesimal area surface in the bulk of the tensor network. From \eqref{eq:diff_ee_icmera} a relation between an AdS geometry and the leading term in \eqref{eq:flm_operator} can be established. To this end, we note that the Ryu-Takayanagi formula for the half space in an asymptotically $(d +2)$-dimensional AdS geometry with a radial direction denoted by $u$:
\footnote{For simplicity, the AdS radius has been fixed to unity and $G_{uu} \to {\text{constant}}$ as $u \to 0$.} 
\begin{align}
 ds^2 = G_{uu}\, du^2 + \frac{e^{2u}}{\epsilon^2}\, d\vx_{d}^2 + G_{tt}dt^2\, ,
\end{align}
is given by \cite{Nozaki:2012zj} 
\begin{align}
 \label{eq:dee_rt}
 d\,S_A=\frac{1}{4 G^{(d+2)}_N}\cdot  d\mathbb{A}_{\text{RT}} \, ,
\end{align}
where
\begin{align}
 d\mathbb{A}_{\text{RT}} \equiv \frac{|A_\perp|}{\epsilon^{d-1}}\, \sqrt{G_{uu}}\,  e^{u(d-1)}\, du
\label{eq:darea_rt}
\end{align}
is an infinitesimal area in the bulk geometry. A direct comparison between \eqref{eq:darea_cmera} and \eqref{eq:darea_rt} suggests that a semiclassical bulk geometry may be inferred from the variational parameters of the tensor network that account for the leading term in \eqref{eq:largeN_ee_chi}. 
Taking this into account, we can read off the $\widehat{\text{Area}}$ operator in icMERA. Recalling \eqref{eq:fidelity-R2}, the differential $d \mathbb{A}_{\text{TN}}$ that can be identified with the RT area surface is related to the expectation value of  $\hat K^{(0)}(u)$ in \eqref{eq:entangler-ON-G-RT}, \emph{i.e.},
\begin{align}
d \mathbb{A}_{\text{TN}} \sim \expval{\hat K^{(0)}(u)^2}_\Omega^{1/2}
 \, .
\label{eq:darea_icmera}
\end{align}

In addition to this result, the remarkable consequence here is that our icMERA analysis allows us to go beyond this semiclassical term in the large $N$ limit by providing an elegant holographic interpretation to the subleading quantum corrections $S_{\textrm q}^{\text{TN}}\equiv S^{(1)}_{\chi\, (\zeta)}$ in terms of the expectation value of the non-Gaussian part of the disentangler as
\begin{align}
S_{\textrm q}^{\text{TN}} \sim \expval{\hat B_N(u)^2}_\Omega^{1/2}
 \, .
\label{eq:qcorr_icmera}
\end{align}

When comparing with \eqref{eq:flm}, the full expression for the first quantum correction to the holographic entropy reads \cite{Faulkner:2013ana}
\begin{align}
S_{\textrm q}=S_{\text{ bulk-ent}} + \frac{\delta\, \mathbb{A}_{\text{RT}}}{4\, G_N} + \expval{\Delta S_{\text{Wald-like}}}+S_{\text{counterterms}}
\, .
\label{eq:ee_q_maldacorr}
\end{align}
The first term is the aforementioned bulk entanglement. The second one represents order $\cO(1)$ fluctuations in the area due to one loop graviton effects that shift the background bulk metric. The third contribution amounts to the quantum expectation value of the formal expression of the Wald-like entropy. Finally, the last term is introduced in order to obtain  a finite result.

Here we hypothesize that $S_{\textrm q}^{\text{TN}}$ is capturing the second term in \eqref{eq:ee_q_maldacorr}, {i.e.},
\begin{align}
S_{\textrm q}^{\text{TN}} = \frac{\delta\, \mathbb{A}_{\text{TN}}}{4\, G_N} = \frac{\expval{\widehat{\delta \mathbb{A}}_{\text{TN}}} }{4 G_N}\sim \expval{\hat B_N(u)^2}_\Omega^{1/2}
 \, .
\label{eq:entropy_qcorr_icmera}
\end{align}
The reason for this is the following. If we consider small perturbations on the bulk metric and perform an expansion over the background, each one of the terms in this expansion have fluctuations and must therefore be related to some operators. In holography, these corrections to the bulk metric must induce corrections on the entanglement entropy. As we have shown, the entanglement entropy is closely related to geometric quantities $\text{Area}(\gamma_A)$ and $\delta \mathbb{A}_{\text{RT}}$ through the variational bond dimension of the icMERA circuit. Precisely, this bond dimension fulfills the requirements exposed in \cite{Bao:2018pvs, Bao:2019fpq}. In these works, it has been established that for a tensor network representing a geometric state in the AdS/CFT, the bond dimension must be determined by the areas of a corresponding extremal surface in the bulk --in cMERA language, the leading term in the cMERA bond dimension must give account of the term ${\text{Area}}/4 G_N$ --, while the subleading terms in the bond dimension must give account of the fluctuations in the areas of those extremal surfaces. In addition, different parts of the icMERA bond dimension are naturally related to fluctuations of operators that correspond to different parts of the icMERA entangler operator. 

This argument, together with the fact that in our derivation from the icMERA circuit, no explicit assumption has been made neither to field excitations in the bulk nor Wald-like entropy expressions, incline us to interpret $S_{\textrm q}^{\text{TN}}$ as   $\delta\, \mathbb{A}_{\text{RT}}$.

\section{Conclusions and Prospects}
\label{sec:conclusions}
We have approached the connection between holography and tensor networks by computing the entanglement entropy of icMERA states corresponding to strongly interacting field theories in the large $N$ limit. Using tensor network \emph{technology} (i.e, quantum information properties of the boundary state), we have derived holographic spatial geometries consistent with the structure of quantum entanglement in the field theory state. Our results suggest how, from the large $N$ limit of icMERA circuits, smooth geometrical descriptions emerge naturally, which represents a crucial feature of the holographic duality. Our holographic interpretation of these tensor networks imply that any two choices of the non-Gaussian part of the circuit generator yield results for the entanglement that differ only at subleading orders in $1/G_N$; that is to say, at the structure of the quantum corrections in the bulk. The fact that the large $N$ part of the entropy can be always related to the leading area term of the holographic calculation suggests a non trivial connection between holography and cMERA. 
Our analysis significantly advances the state of the art of the established field of holography and tensor networks mentioned in the introduction. Namely, our proposal presents a concrete setting in which, one may trace in detail, the emergence of dual geometric descriptions from the entanglement features of states corresponding to a large number of strongly interacting quantum fields.

It is worth to finish by mentioning several prospects for future investigations. First, our results suggest that a bond dimension defined through the Fisher metric only includes field theory data related to the two point functions of the theory. It is thus tempting to find  generalized bond dimension definitions that go beyond the Fisher metric that are able to include corrections due to higher order point functions and then analize their putative holographic interpretations. Secondly, it is a matter for a future investigation to compare our results for the entanglement entropy of an optimized icMERA circuit in the $\lambda\, \phi^4$ theory with those obtained in \cite{Iso:2021vrk,Iso:2021rop}.

\section*{Acknowledgments}
We thank Satoshi Iso, Takato Mori and Katsuta Sakai for very fruitful discussions. The work of JJFM is supported by Universidad de Murcia-Plan Propio Postdoctoral, the Spanish Ministerio de Econom\'ia y Competitividad and CARM Fundaci\'on S\'eneca under grants FIS2015-28521 and 21257/PI/19. JMV is funded by Ministerio de Ciencia, Innovaci\'on y Universidades PGC2018-097328-B-100 and Programa de Excelencia de la Fundaci\'on S\'eneca Regi\'on de Murcia 19882/GERM/15. 

\appendix 

\section{Explicit evaluation of quantum corrections}
\label{app:appendix}
In  icMERA circuits with non-Gaussian entanglers of the type $\pi\, \phi^2$ we have to evaluate the integral over momenta
\begin{align}
\bar \chi_{2}(u)&= \int_{\vp\, \vq}\, c(\vp e^{-u},\vq e^{-u};u)^2
\, ,
\end{align}
where the variational parameter $g_B(u)$ is contained in
\begin{align}
c(\vp e^{-u},\vq e^{-u};u)=g_{B}(u)\cdot \Gamma\left(|\vp +\vq| e^{-u}/\Lambda\right)\, \Gamma\left(p e^{-u}/\Lambda\right)\, \Gamma\left(q e^{-u}/\Lambda\right)\, .
\end{align}
As a result, this quantity factorizes into $g_B(u)^2$ and momentum integrals, such that
\begin{align}
 \bar \chi_{2}(u)&= g_{B}(u)^2 \cdot I_2(\Lambda e^{u})
 \end{align}

with
\begin{eqnarray}
I_2(\Lambda e^{u}) &=& \int_{\vp,\vq}\, \Gamma\left(|\vp +\vq| e^{-u}/\Lambda\right)\, \Gamma\left(p e^{-u}/\Lambda\right)\, \Gamma\left(q e^{-u}/\Lambda\right)
\, .
\end{eqnarray}

A useful formula for doing this momentum integrals is
\begin{dmath}
f(\vec y) = \int\limits_{\mathbb R^d}\text{d}^d\vec x\, \Theta(a-|\vec x|)\,\Theta(a-|\vec x+\vec y|) = \frac{2\pi }{3} a^d \left(1-\frac{|\vec y|}{2a}\right)^2 \left(\frac{|\vec y|}{2a}+2\right) \Theta\left(1-\frac{|\vec y|}{2a}\right)\, ,
\end{dmath}
where $\Theta$ is the Heaviside step function. With this we note that
\begin{dmath}
I_{2}(\Lambda e^{u})= \int_{\vp}\, \Gamma\left(p e^{-u}/\Lambda\right)\, \int_{\vq}\, \Theta(\Lambda e^{u} - |\vp +\vq|) \cdot \Theta(\Lambda e^{u} - q)
=  \frac{2 \pi}{3}\, (\Lambda e^{u})^d\, \int_{\vp}\, \Gamma\left(p e^{-u}/2\, \Lambda\right) \left[ \left(1- \frac{p}{2 \Lambda e^{u}}\right)^2\, \left(2 + \frac{p}{2 \Lambda e^{u}}\right)\right]
 \, .
 \end{dmath}
 
Thus, making $z = p/2 \Lambda\, e^{u}$, we obtain
\begin{align}
I_{2}(\Lambda e^{u}) &= 2^d\, \frac{2 \pi}{3}\, (\Lambda e^{u})^{2d}\, \int_{0}^{1}\, 
dz \left[2 z^{d-1} - 3 z^{d} +z^{d+2}\right]=c_d\, (\Lambda e^{u})^{2d}\, ,
\end{align}

with $c_d = 2^d\, \frac{2 \pi}{3}\, \left[2/d -3/(d+1) + 1/(d+3) \right]$. 

To evaluate the integral $I_3(x)$ in \eqref{eq:def_zeta_u} associated to the $O(N)$ entangler  $B\propto (\bsPi\cdot\bsPhi)( \bsPhi\cdot\bsPhi)$, the same formula can be iteratively applied.

\section{Optimization of the $O(N)$ icMERA circuit}
\label{app:appendix_b}

In \cite{Fernandez-Melgarejo:2020fzw} a detailed description of an icMERA circuit for a single scalar field theory with an entangler of the type $B = \pi \phi^2$ was given. In this appendix we deal with the optimization procedure for an icMERA circuit for the $O(N)$ model with an entangler given by \eqref{eq:entangler-ON-G-RT} and \eqref{eq:entangler-ON-NG}.

The optimization is carried out by minimizing the expectation value of the energy density w.r.t. the icMERA ansatz for a fixed length scale $u$, \emph{i.e.}, $\expval{ \cH }_u = \langle \Psi_u |\, \cH\,  | \Psi_u\rangle\,$. As shown in \cite{Fernandez-Melgarejo:2020fzw} it is rather convenient to take the scale $u_{*} = \log \left(\mu_{*}/\Lambda\right)$, where $\mu_{*}$ is a variational mass parameter that will be defined below, in order to solve the equation for the variational parameters. The optimal values for these parameters are obtained by imposing
\begin{equation}
\label{eq:optimization}
\frac{\delta\, \expval{\cH}_u}{\delta f_{R(T)}(k;u)}=0\, , \qquad  \frac{\delta\, \expval{\cH}_u}{\delta \bar f(\vp,\vq;u)}=0\, ,
\end{equation}
where
\begin{align}
f_{R(T)}(k,u)=\int_0^{u}\, g_{R(T)}(ke^{-\sigma};\sigma)\, d\sigma
\, ,
\end{align}
and
\begin{align}
\bar f(\vp,\vq;u)\equiv f(|\vp + \vq|,p,q;u)=\int_0^{u}\, d\, \sigma\, g(|\vp + \vq|,p,q;\sigma)
\, .
\end{align}

For the \eqref{eq:ON_theory}, the icMERA circuit leaves \cite{IbanezMeier:1991hm}
\begin{dmath}
\expval{\cH} = \expval{\cH}_G  + \beta^2\, \left(\chi_7 + \f12\, m^2\, \chi_2 \right)
+ \lambda\left[4\phi_c\, \beta \chi_3 + \beta^2\left(6I_0^R + (2N-1)I_0^T + 6\phi_c^2 \right)\chi_2\right+ 2\beta^2 \chi_5 + 4\beta^3\, \phi_c \chi_4 + 3\beta^4(\chi_6 + \chi_2^2)]
\, ,
\label{eq:energy_vev}
\end{dmath}
where for clarity, from here in advance, we will drop the $u$-dependence of the variables, understanding that everything is defined at the scale $u_{*}$ mentioned above. As any operator expectation value computed through a non-linear canonical transformation, $\expval{\cH}$ is given by the expectation value obtained by the Gaussian part of the ansatz $\expval{\cH}_G$ plus corrections in powers of $\beta$. The Gaussian part is given by
\begin{dmath}
\expval{\cH}_G = \left(J^R + \f12\, m^2\, I_0^R\right) + (N-1)\left(J^T + \f12\, m^2\, I_0^T \right) + \left(\f12 m^2 \phi_c^2 + \lambda \phi_c^4\right) + 
\lambda \left[3 (I_0^R)^2 + (N^2 - 1)(I_0^T)^2 + 2(N-1) I_0^R I_0^T + 6I_0^R \phi_c^2 + 2(N-1) I_0^T\phi_c^2  \right]
\, .
\end{dmath}
Here
\begin{align}
I_0^{R(T)}&=\f12\, \int_{\vk}\, G_{R(T)}(k)\, ,\quad J^{R(T)} = \f14 \int_\vk \left[ G_{R(T)}(k)^{-1} + k^2\, G_{R(T)}(k)\right]
\, .
\end{align}
we note that in \eqref{eq:energy_vev} it is the term linear in $\beta$, proportional to $\chi_3$, the one assuring an energy improvement compared with the Gaussian expectation value \cite{IbanezMeier:1991hm, Ritschel:1992vr}.

The unbarred $\chi$'s are given by
\begin{align}
\chi_2&=\f12 (N-1)\int_{\vp \vq}\, \bar f(\vp,\vq)^2 \, G_T(p)\, G_T(q)\, ,\\
\chi_3&=\f12 (N-1)\int_{\vp \vq}\, \bar f(\vp,\vq) \, G_T(p)\, G_T(q)\, ,\\
\chi_4&= (N-1)\int_{\vp \vq\vr}\, \bar f(\vp,\vq) \bar f(\vp,\vr)  \bar f(\vq,-\vr) \, G_T(p)\, G_T(q)\, G_T(r)\,\\
\chi_5&= (N-1)\int_{\vp \vq \vr}\, \bar f(\vp,\vq) \bar f(\vp,\vr)  \, G_T(p)\, G_T(q)\, G_T(r)\,\\
\chi_6&= (N-1)\int_{\vp \vq \vr \vk}\, \bar f(\vp,\vq) \bar f(\vp,\vr)  \bar f(\vq,\vk) \bar f(\vr,\vk)\, G_T(p)\, G_T(q)\, G_T(r)\, G_T(k)\, \\
\chi_7 &= \f14 (N-1) \left[\int_{\vp \vq}\, (\vp + \vq)^2\, \bar f(\vp,\vq)^2 \, G_T(p)\, G_T(q) 
\int_{\vp \vq}\, \bar f(\vp,\vq)^2 \, \left(\frac{G_T(p)+G_T(q)}{G_R(|\vp + \vq|)}\right)\right]
\, . 
\end{align}

 Despite the procedure in \eqref{eq:optimization} can be done in full generality, $\phi_c$ has to be fixed, in order for the trial wavefunctions to be consistent with the Rayleigh-Ritz method. In other words, one has to fix $\phi_c$, and minimize with respect to the rest of the variational parameters. In this form, this yields a set of nonlinear coupled equations that must be solved self-consistently and greatly simplify for $\phi_c=0$. The solution for the optimal values for the kernels $G_{R(T)}$ and $\bar f$ are 
\begin{align}
G_{R(T)}(k)=\frac{1}{\sqrt{k^2 + \mu^2_{R(T)}}}\,.
\end{align}
With this we obtain part of the variational parameters of the icMERA ansatz by recalling that in (i)cMERA \cite{Cotler:2016dha}

\begin{align}
f_{R(T)}(k,u_{\text{IR}})=\f12\, \log \frac{G_{R(T)}(k)^{-1}}{M} =\int_0^{u_{\text{IR}}}\, g_{R(T)}(u)\cdot \Gamma(k e^{-u}/\Lambda)\, du
\, .
\end{align} 

On the other hand, the $\bar f(\vp,\vq)$ obeys the Schwinger-Dyson-type integral equation \cite{IbanezMeier:1991hm, Ritschel:1992vr}
\begin{align}
\bar f(\vp,\vq)=F_0(\vp,\vq) \left(1-4\lambda\int_\vr\, \left[\bar f(\vp,\vr) + \bar f(\vq,\vr) \right]\ G_T(r)\right)\, ,
\label{eq:schwinger_dyson}
\end{align}
where
\begin{align}
F_0(\vp,\vq)^{-1}=\left((\vp + \vq)^2 + \mu_{*}^2  + \frac{1}{G_R(|\vp + \vq|)}\, \left[ \frac{1}{G_T(p)} + \frac{1}{G_T(q)}\right]\right)\, .
\end{align}
Here
\begin{align}
\beta = -4\lambda\, \phi_c\, ,
\end{align}   
has been chosen as a way to conveniently normalize $\bar f(\vp,\vq)$ and $\mu_{*}=\mu_{R(T)}|_{\phi_c=0}$.

The solution to this equation can be written as
\begin{align}
\bar f(\vp,\vq) = R(\vp,\vq)\left[\rho(p) + \rho(q)\right]
\, ,
\end{align}
with
\begin{align}
R(\vp,\vq)
=
\frac{G_R(|\vp+\vq|)}{G_R^{-1}(|\vp+\vq|) + G_T^{-1}(p) + G^{-1}(q))}
\ ,
\end{align}
and $\rho(p)$ is determined by a linear integral equation in one variable
\begin{align}
\rho(p)\left[1-\eta(p)\right] = \eta(p) - 4\lambda\int_{\vr}\, R(\vp,\vr) G_T(r) \rho(r)
\, ,
\end{align}
where
\begin{align}
 \eta(p) = - 4\lambda\int_{\vr}\, R(\vp,\vr) G_T(r) 
 \, .
\end{align}

\begin{figure}[!t]
\begin{center}
\includegraphics[height=.15\textheight]{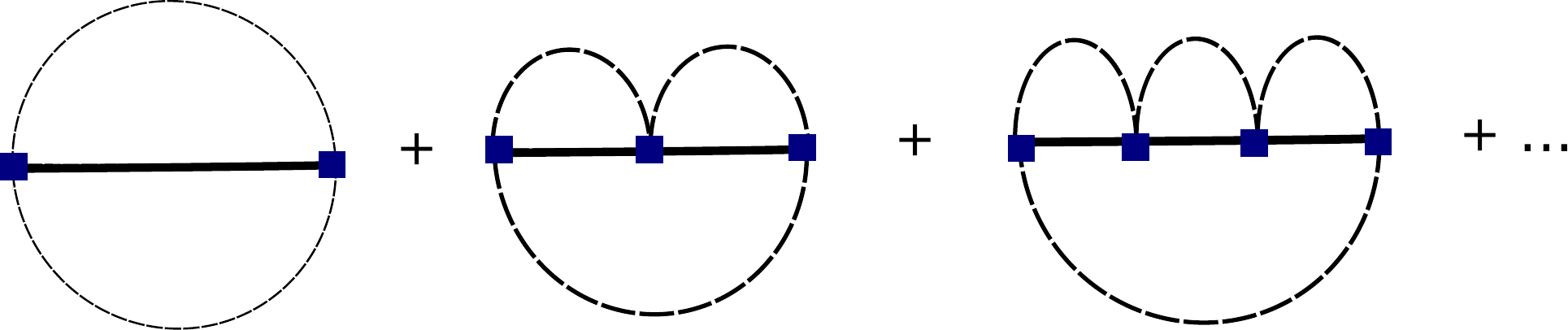}
\caption{\textit{Three contributions in the series of diagrams generated by the
Schwinger-Dyson equation \eqref{eq:schwinger_dyson} contributing to the two-point vertex function . Thick lines represent (R)adial propagators $G_R(p)$ and discontinuous lines represent (T)ransverse propagators $G_T(q)$.}}
\label{fig:feynman_sd}
\end{center}
\end{figure}

An interesting property of this solution is that, in the perturbative regime,  the Schwinger-Dyson iteration in \eqref{eq:schwinger_dyson} implies that the non-Gaussian terms in \eqref{eq:energy_vev} contain an infinite series of contributions to the two-point vertex function from which the
first three graphs are depicted in Figure \ref{fig:feynman_sd}.  This shows that the approximation goes far beyond the Gaussian approximation which only sums up the ``cactus'' graphs.

Finally, we unravel the solution for the variational parameters of the non-Gaussian part of the disentangler by solving for $g_B(u)$:
\begin{align}
	\bar f(\vp,\vq;u_{*})
	&= \int_0^{u_{*}} g(|\vp+\vq| e^{-u},p e^{-u},q e^{-u})du
	=\int_0^{u_{*}} g_B(u)\cdot \Gamma\left(\frac{|\vp+\vq| e^{-u}}{\Lambda}\right) du
	\ .
\end{align}

\section{Other icMERA circuits}
\label{sec:pertholo}

According to the agreement between the entanglement entropy computed through the effective bond dimensions in the icMERA circuits discussed above, and the calculation of the entanglement entropy as the sum of the individual entropies of $N$ effective Gaussian wavefunctionals, we firstly we conclude that our Fisher metric prescription for the effective bond dimension of an icMERA circuit correcly captures the structure of the entropy. 

That said, other non-Gaussian unitary transformations can be chosen to build the circuit for the $O(N)$ vector model. In this section we consider two other $O(N)$ entanglers  (higher order in fields) and study how the variational parameter is related to the coupling. To carry out this analysis we will assume the perturbative regime. For $|\lambda|\ll1$, because the entangler enters the operator algebra that transforms a Gaussian state into the non-Gaussian ground state of the interacting theory, the relation between $\beta$ and the coupling $\lambda$ can be determined \cite{Cotler:2018ehb,Cotler:2018ufx,Fernandez-Melgarejo:2020utg}. Then, upon optimization and dimensional arguments, $\beta$ is fully determined. Interestingly, when calculating the effective bond dimension and taking the large $N$ limit, in both cases the non-Gaussian contributions are subleading with respect to the Gaussian part.

\begin{figure}[!t]
\begin{center}
\includegraphics[height=.10\textheight]{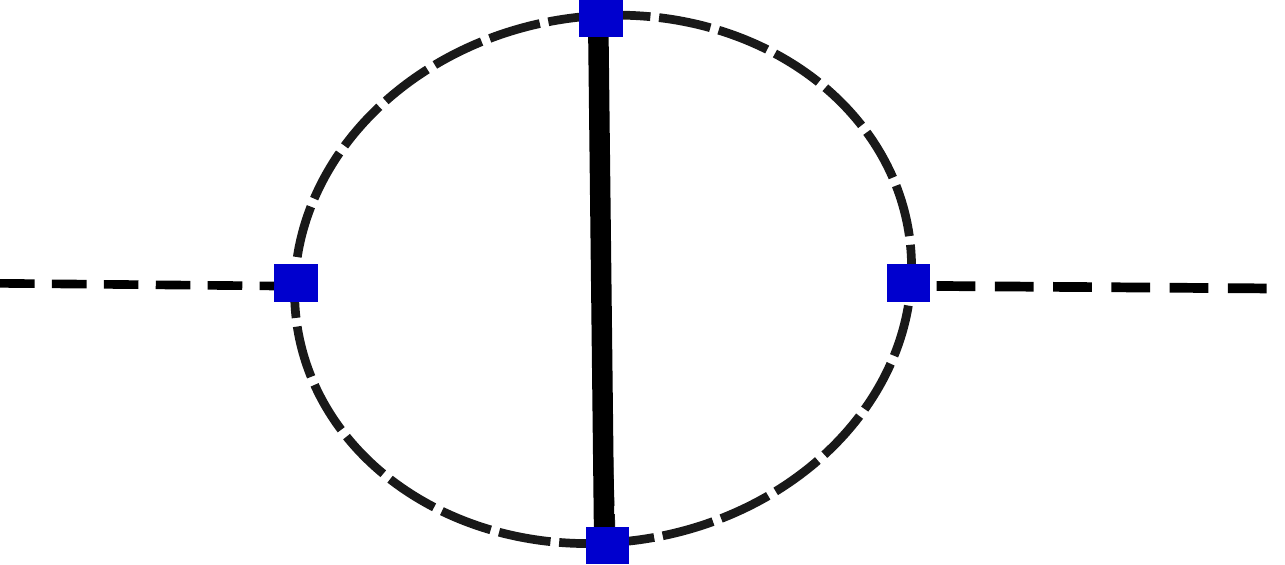}
\caption{\textit{First Feynman diagram contributing, at perturbative level, to the two-point vertex function for the entangler in \eqref{eq:alt_RT_entangler}. Thick lines represent (R)adial propagators $G_R(p)$ and discontinuous lines represent (T)ransverse propagators $G_T(q)$.}}
\label{fig:feynman_piphi4}
\end{center}
\end{figure}

Let us firstly consider the transformation  $\pi_R(\vphi_T\cdot\vphi_T)^2$ given in terms of the non-Gaussian entangler:
\begin{multline}\label{eq:alt_RT_entangler}
\hat B_N(u)
=
\beta \int_{\vq_i} c(\vq_1 e^{-u},\vq_2e^{-u},\vq_3e^{-u},\vq_4e^{-u};u)
\\
\times \pi_R(\vq_1+\vq_2+\vq_3+\vq_4) \ \vec\phi_T(\vq_1)\cdot\vec\phi_T(\vq_2) \ \vec\phi_T(\vq_3)\cdot\vec\phi_T(\vq_4) 
\ ,
\end{multline}
where $c(\vq_1,\vq_2,\vq_3,\vq_4;u) \equiv g_B(u)\, \Gamma(|\sum_i \vq_i|/\Lambda)\Gamma(q_1/\Lambda)\Gamma(q_2/\Lambda)\Gamma(q_3/\Lambda)\Gamma(q_4/\Lambda)$ is a fully symmetric variational parameter. When taking into account the full unitary \eqref{eq:ON-1-icMERA} with $B_N$ given by \eqref{eq:alt_RT_entangler}, the corresponding bond dimension is:
\begin{align}
g_{\text{eff}}(u)^2 
&=\frac{(N-1)}{N}\, g_T(u)^2 + \frac{1}{N}\, g_R(u)^2  + \gamma^2\, e^{-u(d-1)}\,  \left(\frac{(N-1)^4}{N^3}\, \bar X_1(u)^2 + 2\, \frac{(N^2 + 2N)}{N}\, \bar X_2(u)\right)\\ \nonumber
&=\frac{(N-1)}{N}\, g_T(u)^2 + \frac{1}{N}\, g_R(u)^2  +  \gamma^2\, e^{-u(d-1)} \left(\frac{(N-1)^2}{N^3}\, \phi_c^2 + 2\, \frac{(N^2 + 2N)}{N}\, \bar X_2(u)\right)
\ ,
\end{align}
with $\gamma^2 \equiv (\beta^2/16 \Lambda^{(d-1)})(M_R/M_T^4)$ and where, as before, we have defined $\phi_c\equiv\langle \phi_R \rangle$ as
\begin{align}
\phi_c = (N-1)\, \bar X_1(u)\, , 
\qquad
\bar X_1(u) =  \int_{\vp,\vq} c\left(\vp e^{-u},-\vp e^{-u},\vq e^{-u},-\vq e^{-u}\right)
\, ,
\end{align}
and
\begin{align}
\bar X_2(u) &=  \f12\, \int_{\vp_i} c(\vq_1 e^{-u},\vq_2e^{-u},\vq_3e^{-u},\vq_4e^{-u};u)^2
\, .
\end{align}
According to this transformation, the connected non-Gaussian correction to the two-point correlation function is $\propto \lambda^2$. At the perturbative level, this correction is given by the Feynman diagram shown in Figure \ref{fig:feynman_piphi4}. This implies that the optimization equations yield $\beta = (\lambda\, \phi_c)^2$ \cite{Ritschel:1990zs,Ritschel:1992vr}. Then, when taking the large $N$ limit and using \eqref{eq:largeN}, we have that $\gamma^2 = \bar \gamma^2/N^2$ and
\begin{align}
g_{\text{eff}}(u)^2 
&= g_T(u)^2 + \frac{1}{N}\, \left(g_R(u)^2 + 2\, \bar \gamma^2 \, e^{-u(d-1)}\, \bar X_2(u)\right) + \cO(N^{-2})
\, .
\end{align}
Therefore, the effective bond dimension is given by,
\begin{align}
g_{\text{eff}}(u) 
&= g_T(u) + \frac{1}{N}\, \frac{\left(g_R(u)^2 + 2\, \bar \gamma^2\, e^{-u(d-1)}\,  \bar X_2(u) \right)}{2\, g_T(u)} + \cO(N^{-2})
\, .
\end{align}
As in the other cases presented in this paper, the structure of entanglement entropy resulting from this icMERA effective bond dimension, will exhibit a holographic interpretation in terms of the (corrected) Ryu-Takayanagi formula.

A similar result can be obtained for the $O(N)$-symmetry preserving transformation of the form $\vpi_R\cdot\vphi_T (\vphi_T\cdot\vphi_T)^2$ represented by the operator:
\begin{align}
B_N(u)
=
\beta \int_{\vq_i} g(\vq_1,\vq_2,\vq_3,\vq_4,\vq_5,\vq_6;u)\vpi(\vq_1)\cdot\vphi(\vq_2) \ \vphi(\vq_3)\cdot\vphi(\vq_4)\ \vphi(\vq_5)\cdot\vphi(\vq_6)\delta(\tsum_i\vq_i)
\ .
\end{align}

In this case, $\beta\propto\lambda^2$ again, as it can be inferred from the perturbative expansion that relates a Gaussian state with the perturbed ground state of the $\phi^4$ theory \cite{Cotler:2018ehb}.

\newpage

\bibliography{references}
\bibliographystyle{utphys}

\end{document}